\documentclass{article}

\usepackage{arxiv}

\usepackage[utf8]{inputenc} % allow utf-8 input
\usepackage[T1]{fontenc}    % use 8-bit T1 fonts
\usepackage{hyperref}       % hyperlinks
\usepackage{url}            % simple URL typesetting
\usepackage{booktabs}       % professional-quality tables
\usepackage{amsmath,amssymb,amsfonts}
\usepackage{nicefrac}       % compact symbols for 1/2, etc.
\usepackage{microtype}      % microtypography
\usepackage{cleveref}       % smart cross-referencing
\usepackage{graphicx}
\usepackage{natbib}
\usepackage{doi}

\usepackage{algorithmic}
\usepackage{graphicx}
\usepackage{pifont}
\usepackage{textcomp}
\usepackage{multirow}
\usepackage[table]{xcolor}
\usepackage{xspace}
\usepackage{diagbox}
\usepackage{enumitem}
\usepackage[most]{tcolorbox}
\usepackage{tikz}
\newcommand*\emptycirc[1][.8ex]{\tikz\draw (0,0) circle (#1);} 
\newcommand*\halfcirc[1][.8ex]{%
	\begin{tikzpicture}
	\draw[fill] (0,0)-- (90:#1) arc (90:270:#1) -- cycle ;
	\draw (0,0) circle (#1);
	\end{tikzpicture}}
\newcommand*\fullcirc[1][.8ex]{\tikz\fill (0,0) circle (#1);}

\newcommand{\worst}{\cellcolor{red!10}}
\newcommand{\best}{\cellcolor{green!10}}
\usepackage{tablefootnote}
\usepackage[ruled,linesnumbered,noend]{algorithm2e}  % nice ruled header/footer
\SetKwInput{KwIn}{Input}    % customise keyword names
\SetKwInput{KwOut}{Output}
\SetKwComment{Comment}{// }{} % C-style end-of-line comments

\definecolor{gree}{HTML}{1e8449}
\definecolor{orang}{HTML}{b35809}

\newcommand{\chimera}{\textit{Chimera}\xspace}
\newcommand{\dataset}{\textit{ChimeraLog}\xspace}

\newcommand{\cert}{\textit{CERT}\xspace}
\newcommand{\twos}{\textit{TWOS}\xspace}

\usepackage[commandnameprefix=ifneeded]{changes}
\setauthormarkup{}
\definechangesauthor[name={Reviewer 1}, color=black]{R1}
\definechangesauthor[name={Reviewer 2}, color=black]{R2}
\definechangesauthor[name={Reviewer 3}, color=black]{R3}
\definechangesauthor[name={Reviewer 4}, color=black]{R4}
\definechangesauthor[name={Common Reviews}, color=black]{CR}
\usepackage{subcaption}
\usepackage{url}
\usepackage{float}

\newcommand{\paperlogo}{{\includegraphics[height=3.5em]{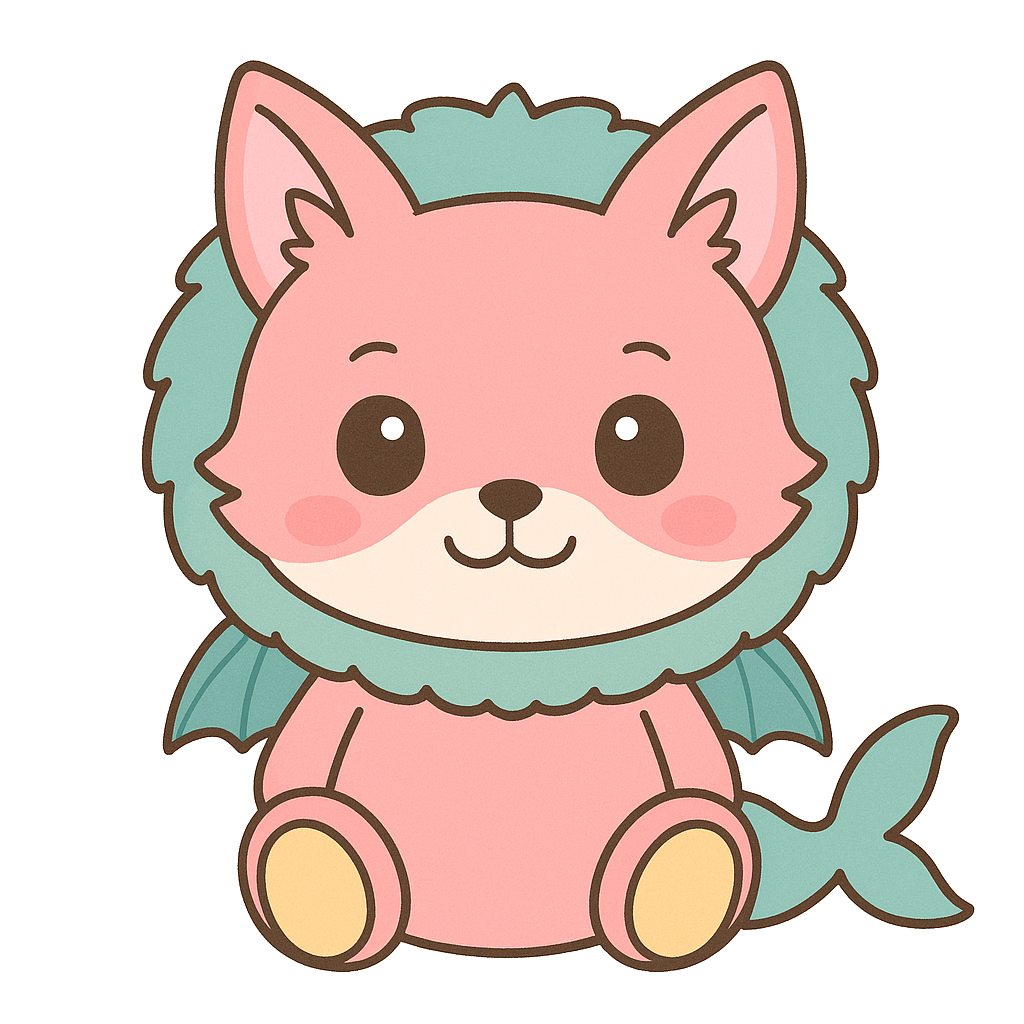}}}

\newcommand{\smallpaperlogo}{\raisebox{-.2em}{\includegraphics[height=2em]{paper-logo.png}}}

\title{
  \begin{center}
    \begin{minipage}[c]{0.12\textwidth}
      \centering
      \paperlogo
    \end{minipage}%
    \hspace{1em}
    \begin{minipage}[c]{0.80\textwidth}
      \centering
      \LARGE Chimera: Harnessing Multi-Agent LLMs for Automatic Insider Threat Simulation
    \end{minipage}
  \end{center}
}

\newif\ifuniqueAffiliation
\uniqueAffiliationtrue

\ifuniqueAffiliation

\author{ \href{https://orcid.org/0000-0002-2888-4499}{\includegraphics[scale=0.08]{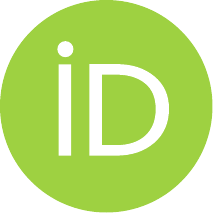}\hspace{1mm}Jiongchi Yu} \\
	Singapore Management University\\
    Singapore, Singapore \\
	\texttt{jcyu.2022@smu.edu.sg} \\
	\And
	\href{https://orcid.org/0000-0002-1288-6502}{\includegraphics[scale=0.08]{orcid.pdf}\hspace{1mm}Xiaofei Xie} \\
	Singapore Management University\\
    Singapore, Singapore \\
	\texttt{xfxie@smu.edu.sg} \\
	\And
	\href{https://orcid.org/0000-0002-8251-1669}{\includegraphics[scale=0.08]{orcid.pdf}\hspace{1mm}Qiang Hu}\thanks{Corresponding Authors.} \\
	Tianjin University\\
	Tianjin, China \\
	\texttt{qianghu@tju.edu.cn} \\
    \And
    {\includegraphics[scale=0.08]{orcid.pdf}\hspace{1mm}Yuhan Ma}\\
	Tianjin University\\
	Tianjin, China \\
	\texttt{mayuhan@tju.edu.cn} \\
    \And
	\href{https://orcid.org/0000-0003-1455-4330}{\includegraphics[scale=0.08]{orcid.pdf}\hspace{1mm}Ziming Zhao$^{*}$}\\
	Zhejiang University\\
	Hangzhou, China \\
	\texttt{zhaoziming@zju.edu.cn} \\
}
\else
\usepackage{authblk}

\newbox{\orcid}\sbox{\orcid}{\includegraphics[scale=0.06]{orcid.pdf}} 
\author[1]{%
	\href{https://orcid.org/0000-0000-0000-0000}{\usebox{\orcid}\hspace{1mm}David S.~Hippocampus\thanks{\texttt{hippo@cs.cranberry-lemon.edu}}}%
}
\author[1,2]{%
	\href{https://orcid.org/0000-0000-0000-0000}{\usebox{\orcid}\hspace{1mm}Elias D.~Striatum\thanks{\texttt{stariate@ee.mount-sheikh.edu}}}%
}
\affil[1]{Department of Computer Science, Cranberry-Lemon University, Pittsburgh, PA 15213}
\affil[2]{Department of Electrical Engineering, Mount-Sheikh University, Santa Narimana, Levand}
\fi

\renewcommand{\shorttitle}{\smallpaperlogo{} Chimera: Harnessing Multi-Agent LLMs for Automatic Insider Threat Simulation}

\hypersetup{
pdftitle={A template for the arxiv style},
pdfsubject={q-bio.NC, q-bio.QM},
pdfauthor={David S.~Hippocampus, Elias D.~Striatum},
pdfkeywords={First keyword, Second keyword, More},
}

\begin{document}
\maketitle

\begin{abstract}
    Insider threats represent a significant and persistent security risk, yet remain difficult to detect in complex enterprise environments, where malicious activities are often concealed within subtle user behaviors. While machine-learning–based insider threat detection (ITD) techniques have shown promising results, their effectiveness is fundamentally constrained by the lack of high-quality and realistic training data. This challenge stems from the highly sensitive nature of enterprise internal data that is rarely accessible and from the limitations of existing datasets, where public datasets are typically small in scale, and synthetic datasets often lack sufficient generalization, rich semantic context, and realistic behavioral patterns.

To address this challenge, we propose \chimera, a large language model (LLM)-based multi-agent framework that automatically simulates both benign and malicious insider activities and monitors comprehensive system logs across diverse enterprise environments. \chimera models each agent as an individual employee with fine-grained roles and incorporates group meetings, pairwise interactions, and self-organized scheduling to capture realistic organizational dynamics. Based on 15 insider attack types abstracted from real-world incidents, we deploy \chimera in three representative data-sensitive organizational scenarios and construct a new dataset, \dataset, for supporting the development and evaluation of ITD methods.

We evaluate \dataset through comprehensive human studies and quantitative analyses, demonstrating its diversity and realism. Experiments with existing ITD methods show that detection performance on \dataset is substantially lower than existing ITD datasets, indicating a more challenging and realistic benchmark. Despite distribution shifts, ITD models trained on \dataset exhibit strong generalization capability, highlighting the practical value of LLM-based multi-agent simulation for advancing ITD.
\end{abstract}

% keywords can be removed

% Introduction
\section{Introduction}

Insider threats refer to security incidents that originate from within an organization and have become a critical concern in modern enterprise environments. Recent studies report that over 50\% of organizations have experienced insider incidents, with 29\% incurring remediation costs exceeding \$1 million dollars~\cite{ponemon2025insider,gurucul2024insider}. Detecting such threats remains particularly challenging as attackers are typically trusted insiders who possess legitimate access to organizational systems. Their malicious activities often blend into normal behavior and manifest in diverse forms, including horizontal propagation such as lateral movement within internal networks, unauthorized data exfiltration, IT sabotage, fraud, and espionage, as well as vertical escalation such as privilege escalation. Well-known incidents illustrate the severity of these threats, such as Edward Snowden's data exfiltration at the NSA~\cite{harding2014snowden} and employee-driven IT sabotage at Tesla in 2018~\cite{agnihotri2024tesla}. Traditional security defenses struggle to identify insider threats due to the legitimate access privileges of insiders, and this challenge is further exacerbated by inadequate monitoring infrastructure. Recent reports show that only 36\% of organizations have deployed comprehensive monitoring systems for proactive ITD~\cite{gurucul2024insider,signpost2025insider}, revealing significant gaps in current enterprise defenses.

\begin{figure}[t]
    \centering
    \includegraphics[width=1\linewidth]{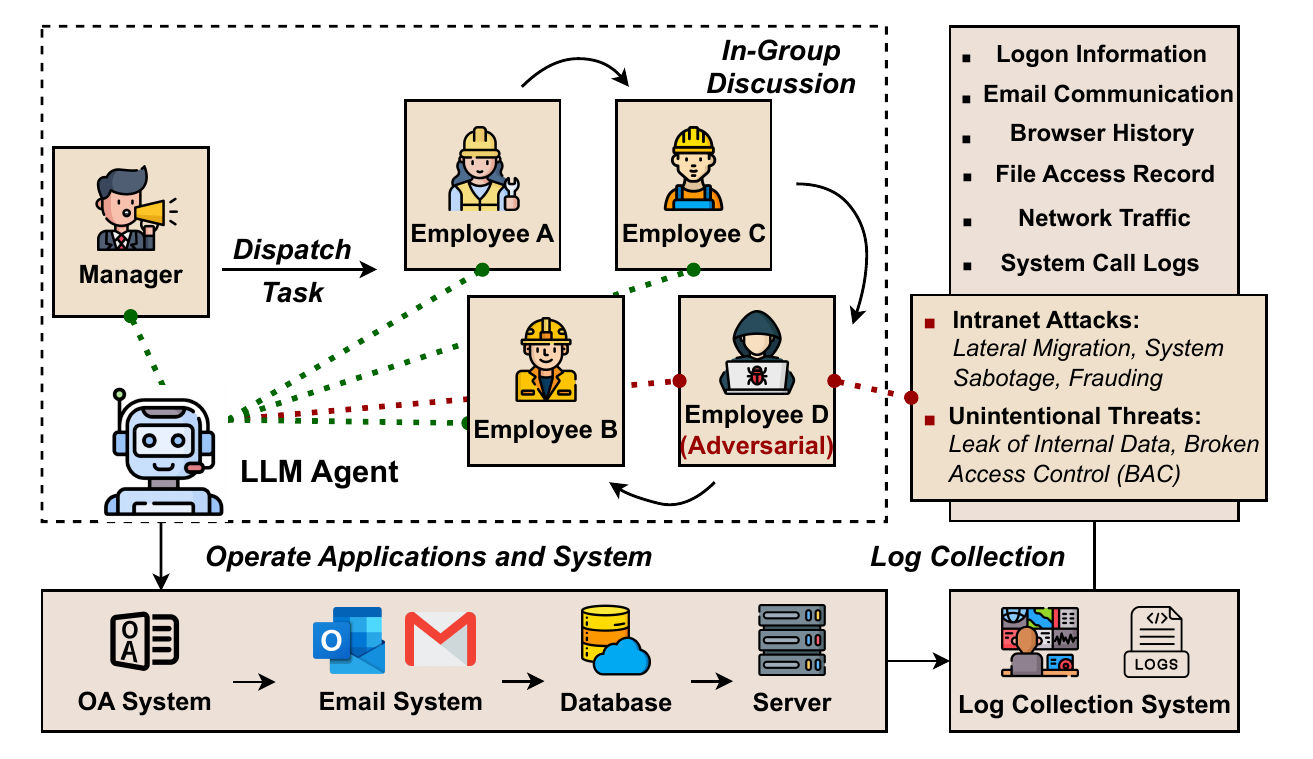}
    \caption{Automation of insider threat simulation.}
    \label{fig:overview}
\end{figure}

To mitigate insider threats, many insider threat detection (ITD) techniques have been proposed, and log-based ITD has emerged as the most promising direction, which analyzes internal activity logs such as authentication records and employee communications for threat identification~\cite{he2024double,hong2022graph,kotb2025novel}. Rule-based approaches typically require substantial manual effort to define detection rules and curate reference data tailored to specific systems. These methods are often ad hoc and constrained to internal deployment due to privacy concerns. They also suffer from high false alarms as systems and user behaviors evolve~\cite{gurucul,erola2022insider}. As a result, frequent manual updates are required, which makes rule-based solutions costly and difficult to generalize.

Machine learning based ITD methods have shown promising detection performance~\cite{janjua2020handling,hong2022graph,kotb2025novel}. However, these approaches rely on large-scale, high-quality, and accurately labeled datasets to distinguish subtle insider threat behaviors from legitimate activities. In practice, acquiring such datasets remains a major obstacle and fundamentally limits the effectiveness and deployability of learning based ITD systems.

Specifically, the lack of high-quality datasets for ITD presents four key challenges: \ding{182} \textit{Privacy Constraints.} Insider activities inherently involve sensitive and proprietary organizational data. Therefore, such data are difficult to share outside the organization for research and analysis purposes, which severely limits the availability of realistic datasets. \ding{183} \textit{Unrealistic Data.} Most publicly available ITD datasets, such as the CERT insider threat dataset~\cite{glasser2013bridging}, are synthetic and lack semantic richness. Both benign behaviors and threat scenarios are manually constructed rather than derived from authentic interactions in real organizational environments~\cite{yuan2021deep,lindauer2014generating}. In addition, these datasets often omit important system-level log modalities that are common in practice, including network traffic and system call logs, which reduces their realism and practical utility. \ding{184} \textit{High Cost.} Collecting and labeling insider threat data from real-world environments is prohibitively expensive due to the scale and complexity of internal activity logs. The cost increases as systems evolve rapidly and as organizations expand their infrastructure and user base. Large enterprises may generate millions of log entries each day, which leads to substantial labeling and maintenance overhead as the organization grows. For example, TWOS~\cite{harilal2017twos}, a dataset collected from human participants in controlled yet realistic settings, is on a small scale due to this high cost, limiting its scenario coverage and practical usage. \ding{185} \textit{Lack of Adaptability.} Enterprise systems are frequently updated, which introduces significant distribution shifts in log data~\cite{han2023anomaly,yu2025cashift}. These shifts can substantially degrade the performance of ITD models trained on outdated datasets. Moreover, insider threat scenarios in existing datasets are often tailored to specific system configurations and do not generalize well. Consequently, datasets require continuous updating and maintenance, which further increases cost and operational complexity.

\textbf{Motivation.}
The aforementioned challenges significantly restrict the preparation of high-quality datasets for ITD, consequently hindering the research and development of effective ITD methods. Without realistic and representative data, existing approaches often suffer from high false positive (FP) rates and poor generalization when deployed across diverse enterprise environments. This gap motivates the need for automated approaches that can generate high-fidelity insider threat datasets while avoiding high cost and privacy risks.

To support practical security practices, an automated ITD data generation framework, as shown in Figure~\ref{fig:overview}, should ideally incorporate the following key attributes: \ding{182} It should enable flexible scenario simulation tailored to domain-specific software and communication protocols. \ding{183} It should realistically model benign user behaviors that reflect genuine organizational activities. \ding{184} It should support adaptive simulation of diverse insider threat behaviors. \ding{185} It should provide comprehensive log collection with accurate labeling to reduce reliance on manual annotation.

Recent advances in large language models (LLMs) and LLM-based agents have demonstrated strong capabilities in simulating human behaviors across domains such as software development~\cite{he2024llm,jin2024llms,yu2025autoempirical}, society systems~\cite{lin2023agentsims,xie2024can}, and automated red teaming~\cite{deng2024pentestgpt,yang2025pentesteval,shen2025pentestagent}. Building on this progress, we propose \chimera, a LLM-based multi-agent framework designed for generalized insider threat simulation. \chimera constructs adaptive enterprise environments and leverages LLM-driven agents to simulate organizational activities in a fully automated manner. Each agent represents an individual organizational member with a specific role, personality, and set of responsibilities. To ensure realistic simulation of both benign behaviors and insider threats, \chimera incorporates several insider-specific mechanisms, including a multi-stage task specification workflow for organizing daily activities, reflective memory that enables agents to maintain behavioral consistency over time, and unrestricted context-rich communication that supports realistic interaction and coordination.

Given a scenario, \chimera automatically generates organizational structures, employee roles, and agent personalities or accepts user-specified configurations. Agents independently plan and execute semantically coherent daily activities such as meetings, email communication, and code execution. Insider attacks are simulated by dedicated attacker agents that follow penetration testing paradigms while continuing routine work to avoid detection. Attacks are guided by abstract attack specifications that allow malicious agents to adapt techniques to the organizational context.

We deploy \chimera in three representative data-sensitive enterprise domains (technology companies, financial institutions, and medical organizations). We simulate a 20-person organization and model 15 insider threat scenarios over a one-month period. This process yields \dataset, which contains approximately 20 billion benign log entries and 5 billion attack log entries across six log modalities. These include application-level logs such as login records, email communication, web browsing, and file operations, as well as system-level logs, including network traffic and system logs. 

To evaluate the effectiveness of \chimera and the quality of \dataset, we conduct a comprehensive evaluation that includes quantitative analyses and human studies. This evaluation assesses dataset realism and the fidelity of \chimera in simulating authentic insider threat scenarios. We further benchmark existing ITD methods to examine their detection performance, focusing on both the challenge posed by \dataset and the generalization of these methods across diverse scenarios and distribution shifts. Our results show that all existing ITD methods experience substantial performance degradation under distribution shifts, underscoring the need for automated simulation frameworks such as \chimera. At the same time, \dataset proves more challenging than existing datasets, while models trained on \dataset demonstrate stronger generalization capability.

% To evaluate the performance of \chimera and the quality of \dataset, we first perform a comprehensive evaluation comprising quantitative analyses and human studies to assess the quality of \dataset. This evaluation rigorously assesses the dataset quality in terms of realism and the effectiveness of \chimera in simulating authentic insider threat scenarios. Additionally, we conduct benchmarking analyses on existing ITD methods to evaluate their detection performance, investigating both the challenge presented by \dataset and the generalization capability of these ITD approaches across diverse scenarios and distributional shifts. We observe that all existing ITD methods suffer significantly under distribution shifts, highlighting the necessity for automated simulation frameworks like \chimera. Meanwhile, we find that \dataset is more challenging for ITD, and models trained on \dataset exhibit stronger generalization than existing datasets.

\textbf{Contributions.} We summarize our contributions as follows:

\begin{itemize}[leftmargin=*]

    \item We design and develop a novel LLM-based multi-agent framework, named \chimera, to simulate the user behavior of enterprise employees and insider threats. \chimera supports diverse enterprise scenarios and organizational roles, enabling the automated generation of realistic and diverse log events without manual behavior scripting.

    \item Using \chimera, we construct a new dataset named \dataset. \dataset covers 15 insider attack scenarios derived from real-world cases and includes six complementary log modalities. It contains approximately 25 billion log entries with fine-grained labels, representing 160 hours of simulated enterprise activity.

    \item We conduct extensive evaluations to assess the quality of \dataset and to benchmark existing ITD methods on this dataset. Results from quantitative analyses and human studies show that \dataset exhibits higher complexity and realism comparable to real-world ITD datasets.
    
    \item We provide empirical insights from deploying \chimera in realistic enterprise settings. Our findings demonstrate the potential of LLM-based multi-agent frameworks for fully automated data generation and analysis in security domains. To support reproducibility and future research, we publicly release the code and dataset on our website~\cite{website}

\end{itemize}

% Background
\section{Background and Related Works}

\subsection{Threat Model}

\textbf{System Model.} We consider an enterprise environment in which employees operate under well-defined organizational roles, such as developers, analysts, and administrators. In \chimera, each employee is modeled as an LLM-powered agent instantiated within an isolated virtual environment and governed by role-based access control policies. This design allows agents to exhibit individualized behaviors while remaining constrained by realistic permission boundaries.

To support comprehensive activity monitoring, \chimera collects logs from six complementary modalities. Four modalities operate at the application layer, including login activity, email communication, web browsing, and file operations. Two additional modalities operate at the system layer, namely network traffic~\cite{zhao2025towards} and system calls~\cite{zhao2023cmd}. Both benign and malicious behaviors are represented as unified event sequences that can be mapped to multi-stage enterprise workflows~\cite{zhao2024tpe}. Each simulated malicious action is further annotated with MITRE ATT\&CK tactics and techniques (TTPs), which enables traceability between simulated behaviors and real-world adversarial patterns. In addition to individual activities, \chimera models essential organizational dynamics such as daily task scheduling, collaborative meetings, and peer-to-peer communication. These mechanisms preserve the temporal consistency and contextual coherence of the generated activity logs.

\textbf{Attacker Capabilities and Assumptions.}  
As summarized in Table~\ref{tab:dataset}, \chimera models three insider archetypes with distinct intentions, privileges, and stealth capabilities. 
These adversarial agents are instantiated within the simulation to emulate realistic and context-aware attack chains.  
\ding{182} \textit{Malicious insiders} are legitimate employees who intentionally abuse their authorized access to harm the organization. Their objectives may include intellectual property (IP) theft, financial fraud, or operational sabotage. They often leverage their contextual knowledge of internal workflows and privileges to escalate access, exfiltrate data covertly, or manipulate systems while maintaining the appearance of normal behavior.
\ding{183} \textit{Masqueraders} are external adversaries who gain unauthorized access by compromising internal credentials, commonly via phishing or credential leakage. Although they lack long-term familiarity with the environment, they can impersonate employees to steal sensitive information, disrupt services, or act as covert conduits for third-party infiltration. Their success often depends on mimicking typical usage patterns to evade detection mechanisms.
\ding{184} \textit{Unintentional insiders} are well-meaning employees whose negligent actions, such as misconfigurations, weak passwords, or falling victim to phishing, lead to unintended security breaches. While lacking malicious intent, they may still cause significant damage, including inadvertent data exposure or the unintentional escalation of adversarial access. These behaviors are modeled as stochastic deviations from normal task sequences, allowing \chimera to reproduce the noise and human patterns observed in real logs.

In real-world incidents~\cite{doj2025laatsch,wilkens2021multi}, insiders often execute multi-phase operations with long dwell periods before detection. Accordingly, \chimera incorporates three hybrid, multi-stage attacks based on real-world cases~\cite{tavani2014trust,departmentofjustice2003review}. In this way, we aim to challenge ITD systems in realistic and complex ways. The resulting logs contain temporally correlated traces across multiple modalities (emails, file access, network flows), mirroring the subtle footprint of persistent insider campaigns.  
For example, an insider may gradually escalate privileges through legitimate administrative requests and later exfiltrate design files via a cloud service, appearing benign except for minor anomalies. It would be challenging for detection models to identify such hidden threats.

% In real-world cases~\cite{doj2025laatsch,wilkens2021multi}, most insider attack incidents involve multiple combined attack stages, and certain attacks may involve long incubation periods during which malicious activities remain undetected, effectively catching system maintainers off guard. Therefore, we include three hybrid attacks based on real-world impactful insider threat cases~\cite{tavani2014trust,departmentofjustice2003review,doj2020kriuchkov} for analysis. 

\textbf{Trust Relationships.}  
Our threat model assumes that the underlying infrastructure, such as the host operating systems, containerization platform, and logging mechanisms, is fully trusted and uncompromised. These components serve solely as a reliable substrate for executing simulation logic and capturing behavioral traces without interference or tampering.
In contrast, all LLM agents representing employees are considered untrusted, including those modeling malicious or colluding insiders. These agents may attempt to exploit system permissions, circumvent access controls, or inject adversarial behavior into collaborative workflows. Notably, even agents simulating benign employees may act as inadvertent threat vectors by processing adversarial content, such as phishing emails or misleading peer communications. \added[id=R3]{To preserve both the realism and integrity of enterprise operations, \chimera enforces agent executions within containers and isolated networks, ensuring that no simulated attack can affect the host or external systems, in alignment with the ethical safeguards.}

\subsection{Log-Based Insider Threat Detection}

Insider threats typically refer to security risks that originate within the trusted boundary of an organization~\cite{homoliak2019insight}. Insiders have intimate knowledge of the organization’s systems and normal procedures, which can enable them to carry out malicious activities in ways that are difficult to distinguish from routine actions. Previous research~\cite{homoliak2019insight,mazzarolo2019insider} categorizes insider threats into broad types based on the perpetrator's relationship to the organization and intent. For example, insiders can be divided into masqueraders and traitors. A masquerader is an outside actor or unauthorized user who manages to gain insider credentials and impersonate a legitimate user, while a traitor abuses his or her privileges to perform malicious acts. In our work, we consider both types of attackers but include unintentional insiders, who inadvertently cause harm without malicious intent.

\begin{table}[t]
\centering
\caption{Comparison of ITD datasets. \textit{App.}, \textit{Net.}, and \textit{Sys.} denote the availability of application logs, network traffic, and system logs, respectively.}
\label{tab:dataset}
\resizebox{1\linewidth}{!}{
\begin{tabular}{l|cccc|cc}
\hline
\textbf{Dataset} & \textbf{App.} & \textbf{Net.} & \textbf{Sys.} & \textbf{Personality} & \textbf{Size} & \textbf{Attack Types} \\ \hline
CERT r6.2~\cite{glasser2013bridging}        &  $\checkmark$ &   &   &  $\checkmark$ & \fullcirc & \halfcirc \\ \hline
TWOS~\cite{harilal2017twos}                 &  $\checkmark$ &  $\checkmark$ &   &  $\checkmark$ & \halfcirc & \emptycirc \\ \hline
CIC-IDS 2017/2018~\cite{sharafaldin2018toward}   &   &  $\checkmark$ &   &   & \fullcirc & \halfcirc \\ \hline
LANL 2017~\cite{kent2015cybersecurity}      &   &  $\checkmark$ &  $\checkmark$ &   & \fullcirc & \emptycirc \\ \hline
WUIL~\cite{wuildata}                        &  $\checkmark$ &   &  $\checkmark$ &  & \halfcirc &  \halfcirc \\ \hline
CPTC 2018~\cite{cptc2018}                   &  $\checkmark$ &  $\checkmark$ &   &   & \halfcirc & \halfcirc \\ \hline
OpTC~\cite{optc2020}                        &  $\checkmark$ &  $\checkmark$ &  $\checkmark$ &   & \fullcirc & \halfcirc \\ \hline
\textbf{Chimera (Ours)}                     &  $\checkmark$ &  $\checkmark$ &  $\checkmark$ &  $\checkmark$ & \fullcirc & \fullcirc \\ \hline
\end{tabular}
}
\end{table}

\textbf{Existing ITD Datasets.} Due to privacy and legal constraints, internal enterprise activity data is rarely released publicly. Consequently, most existing ITD datasets are either synthetically generated~\cite{glasser2013bridging,sharafaldin2018toward} or collected through controlled environments with limited realism~\cite{harilal2017twos}. Table~\ref{tab:dataset} compares representative ITD datasets across several key dimensions. Specifically, 
\textit{CERT}~\cite{glasser2013bridging} is one of the most widely used synthetic ITD datasets. It includes application-layer logs such as logon records, file accesses, and email communications, generated for over 100 simulated users over several months. However, the log data lacks semantic information, and its behaviors are rule-based and repetitive, which limits realism. \textit{TWOS}~\cite{harilal2017twos} captures human-generated activity from a controlled five-day red-team/blue-team competition involving university students. While its logs exhibit greater behavioral authenticity, TWOS is constrained in both scope and scale. \textit{CIC-IDS 2017}~\cite{sharafaldin2018toward} focuses on external intrusion detection rather than insider threats. It simulates network traffic using predefined benign profiles and inserts attacks such as DDoS, brute-force, and botnets. However, it lacks application-level logs and any notion of user roles or personalities. \textit{LANL 2017}~\cite{kent2015cybersecurity} provides a rich set of real internal authentication and network flow logs over 58 days. Despite its scale and realism, it only covers identity management and system-level behaviors, with no application semantics or attack annotations. \textit{WUIL}~\cite{wuildata} (Windows User Interaction Logs) consists of real Windows GUI usage traces (e.g., mouse events, file access, registry changes). It reflects fine-grained user activity but lacks labeled threat behaviors. \textit{CPTC 2018}~\cite{cptc2018} captures red-team/blue-team activity from a collegiate penetration testing competition. While it contains real attacker actions and offers some application and network logs, the data is noisy, fragmented, and lacks continuity of benign behavioral patterns. \textit{OpTC}~\cite{optc2020} is a DARPA-backed dataset collected from simulated real-world enterprise environments with embedded red team attacks. It includes comprehensive logs across multiple modalities and timeframes, but still lacks semantic realism in user behaviors.

\textbf{ITD Methods.} Existing work has explored a wide range of machine learning-based approaches to detect insider threats in large organizations. Early approaches applied traditional classifiers on engineered features from system logs. For example, Support Vector Machine (SVM)~\cite{janjua2020handling} classifiers have been trained on user activity statistics to distinguish benign from malicious profiles. These methods can achieve very high accuracy on balanced datasets, and they benefit from interpretability and strong performance with limited data.

Deep neural networks have also been adopted. Convolutional Neural Networks (CNNs)~\cite{hong2022graph} have been applied by converting user behavior sequences into ``image-like'' matrices and using convolution and pooling to extract spatiotemporal features. Graph-based models, such as Graph Convolutional Networks (GCNs)~\cite{hong2022graph}, treat an organization’s user interactions as a graph, propagating each user's profile and activity features along edges to capture relational and community structure. More recently, hybrid deep models have emerged. For example, Deep Synthesis-based Insider Intrusion Detection (DS-IID) model~\cite{kotb2025novel} uses deep feature synthesis to automatically generate rich user profiles from event logs, and then applies a binary deep learning classifier to detect insiders.

Recently, the rise of LLMs has brought new opportunities for ITD. Audit-LLM~\cite{song2024audit} proposes a multi-agent collaboration framework, where different LLM agents are assigned specialized roles to jointly analyze large-scale organizational logs. Similarly, LogGPT~\cite{qi2023loggpt} leverages the powerful language modeling capability of LLMs to process log sequences as natural language, demonstrating strong performance in anomaly detection tasks under zero-shot settings. RedChronos~\cite{li2025redchronos} introduces a production-level LLM-based log analysis system, incorporating query-aware voting and semantic-expansion algorithms to enhance detection accuracy and automation in real-world SOC operations. While these studies employ LLMs for log analysis or limited log synthesis, they do not simulate end-to-end organizational operations with multi-modal enterprise logs as part of an autonomous simulation framework.
 
\subsection{LLM-Based Multi-Agent Systems}

Multi‑agent systems have emerged as a transformative paradigm in both cybersecurity and software engineering, offering decentralized, collaborative, and adaptive solutions to complex challenges. Comprising multiple autonomous agents that interact within shared simulated environments, MAS enable scalable and resilient architectures capable of handling tasks that are difficult for monolithic systems to manage effectively. Established frameworks such as CAMEL~\cite{li2023camel}, AutoGen~\cite{wu2023autogen}, and MetaGPT~\cite{hong2023metagpt} exemplify this trend. More recent open‑source and simulation-focused systems like AgentSims~\cite{lin2023agentsims} provide GUI‑driven sandboxes for evaluating LLM agents in custom social and planning tasks, while Alympics~\cite{mao2023alympics} applies game‑theoretic settings to probe strategic decision‑making by LLM agents. Specialized simulation platforms such as BotSim~\cite{qiao2025botsim} model malicious social botnets using LLM‑powered bots to emulate coordinated misinformation campaigns, and social‑network simulation systems with LLM‑empowered agents explore emergent social behavior in networked settings. MAS can also realistically simulate human user behavior for evaluation and modeling~\cite{wang2025user}. In medical domains, MedSentry~\cite{chen2025medsentry} explores safety vulnerabilities in LLM‑based multi‑agent architectures under adversarial prompts. Although existing MAS frameworks support LLM-driven collaboration and behavioral simulation, none have been designed to reproduce the full spectrum of enterprise operations or to synthesize temporally aligned, multi-modal security logs.

% Design of the framework
\begin{figure*}[ht]
    \centering
    \includegraphics[width=1\linewidth]{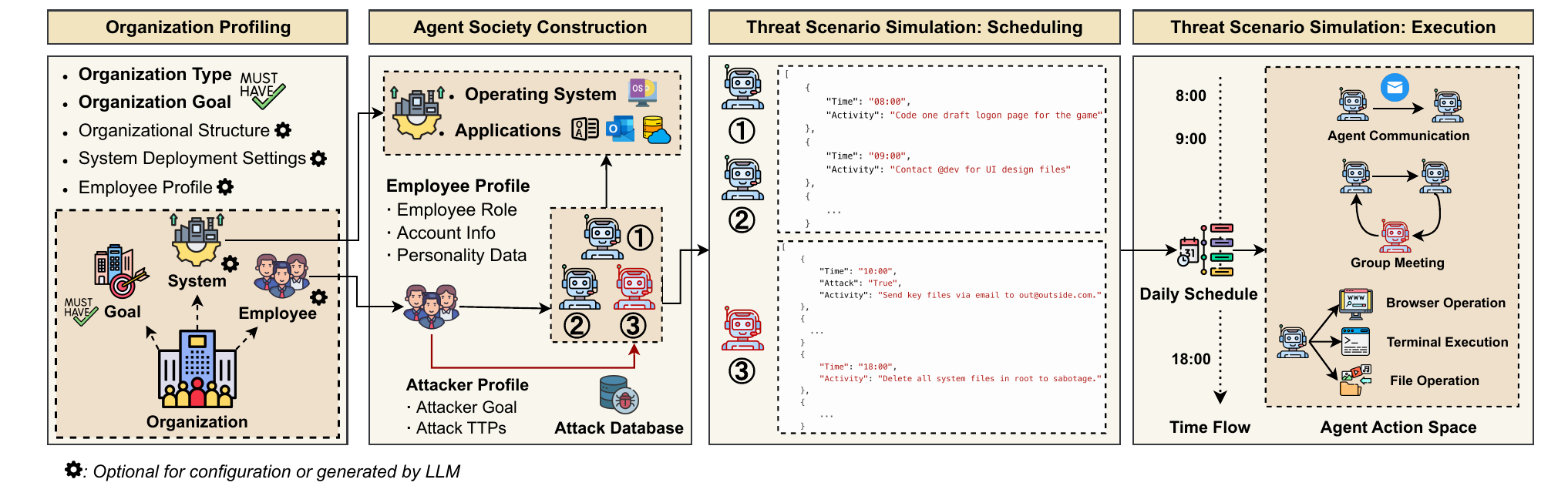}
    \caption{The workflow of \chimera for automated insider threat simulation.}
    \label{fig:workflow}
\end{figure*}

\section{Design of Chimera}

\subsection{Overview} 

The goal of \chimera is to integrate LLM-based multi-agent systems to simulate organizational operations, such as the daily activities of employees in a large enterprise. Each agent represents an individual employee and performs benign or potentially malicious actions within the simulated environment. As illustrated in Figure~\ref{fig:workflow}, \chimera operates through three main phases: \textit{Organization Profiling}, \textit{Agent Society Construction}, and \textit{Threat Scenario Simulation}.

Algorithm~\ref{alg:chimera} summarizes the workflow. Specifically, \ding{182} given the organization type (e.g., game company) and goal (e.g., develop a new game), \chimera either accepts a user-provided configuration or automatically generates an organizational profile, including system deployments, employee numbers, and assigned roles within the organization under simulation (OUS).
\ding{183} Next, \chimera constructs an agent society in which each agent is instantiated according to an assigned organizational role and personality profile. A subset of agents is designated as adversarial insiders and configured with specific attack objectives.
\ding{184} Finally, \chimera simulates daily organizational activities. Benign agents generate task schedules aligned with normal work objectives, while adversarial agents embed malicious actions into otherwise legitimate routines. As agents collaborate to achieve organizational goals such as software development, \chimera continuously records application-level and system-level logs.

{\small
\begin{algorithm}[t]
  \caption{Workflow of \chimera}
  \label{alg:chimera}
  \KwIn{%
    Scenario Configuration $X=(E, R, S, G, T)$: 
    Employees $E$, Roles of Employees $R$, System Environments $S$, Organization Goal $G$, and Simulation Duration $T$.
  }
  \KwOut{%
    Application\,/\,System-Level Logs $L$.
  }

  \tcp{Phase 1: Organization Profiling}
  \If{\textsc{IsMissing}($X$)}{ 
      $X \leftarrow$ \textsc{GenerateOrgProfile\_LLM}($G$)\;
  }
  \textsc{SetupSystemEnv}($X$)\;

  \tcp{Phase 2: Agent Society Construction}
  $A \leftarrow \varnothing $ \tcp*[h]{Initialize agents}\;
  \ForEach{employee $e \in E$}{
      $a \leftarrow$ \textsc{CreateAgentBundle}($e$)\;
      \textsc{AssignProfile}($a$, $r(e)$)\;
      \textsc{EquipTools}($a$, \{terminal, browser, file‐ops\})\;
      \If{$e$ is Adversarial}{%
        \textsc{AssignAttackObjective}($a$, $G$)\;
      }
      $A \leftarrow A \cup \{a\}$\;
  }

  \tcp{Phase 3: Threat Scenario Simulation}
  \For(\tcp*[f]{day-level time loop}){$t=1$ \KwTo $T$}{
      \ForEach{agent $a \in A$}{
          \textsc{GenerateDailySchedule}($a$, $t$)\;
          \If{$a$ is adversarial}{%
              \textsc{UpdateAttackSchedule}($a$, $t$, $P$)\;
          }
      }
      \ForEach{timeslot $\tau$ on day}{
          \ForEach{agent $a \in A$ \KwSty{parallel}}{
              \textsc{ExecuteTasksOrAttack}($a$, $\tau$)\;
          }
          \textsc{UpdateSchedulesAfterComms}($A$, $\tau$)\;
      }
  }
  \tcp{Phase 4: Unified Logging}
  $L \leftarrow \textsc{CollectLogs}(A, P)$\;
  \Return $\langle L, P \rangle$\;
\end{algorithm}
}

\subsection{Organization Profiling}

To realistically simulate organizational operations, \chimera first instantiates both the organizational structure and the system settings. The organizational structure specifies employee counts, role assignments, and the high-level business objective (e.g., a game studio developing a new game or a financial firm running a trading project), which directly shapes application-level behaviours and the semantic content of logs. System settings capture platform details and deployed services (OS variants, Office Collaboration stacks, mail servers, browsers, etc.), since even minor configuration or version differences can materially change system-level artifacts~\cite{dragoi2022anoshift,yu2025cashift}.

As exact enterprise configurations are often unavailable, \chimera supports two operating modes: \ding{182} accept user-provided profiles for scenario-specific fidelity, or \ding{183} automatically synthesize plausible organizational profiles using structured templates and LLM-guided generation. For reproducibility and controlled experiments, each simulated organization is deployed as a standardized container pre-provisioned with representative enterprise services such as email servers. To ensure realistic simulations, \chimera can construct organizational settings grounded in established models of multi‑agent coordination following prior works~\cite{hong2023metagpt} and scale the number of employees to align with the intended organizational goals.

\subsection{Agent Society Construction}

With the organizational configuration and environment established, \chimera proceeds to construct an LLM-based agent society, where each employee in the organization is represented by an autonomous agent. Building on prior work~\cite{li2023camel,yang2024oasis}, we employ multiple collaborative LLM agents per employee, forming an ``agent bundle'' for greater behavioral realism. Each bundle comprises functional agents equipped with specialized tools, including a \textit{user} agent (responsible for task planning) and an \textit{assistant} agent (responsible for executing actions using tools such as the terminal, browser, and file operation utilities). These tools are configured with fine-grained prompts derived from open-sourced frameworks (e.g., CAMEL~\cite{li2023camel}) to ensure accurate and consistent interactions with the system environment.

Each agent bundle is initialized with employee metadata (e.g., name, role, department) and system identifiers (e.g., container ID), forming a unified employee profile. To simulate adversarial behavior, \chimera designates agents as insiders who execute threat activities within the system based on the attack configuration. These adversarial agents are assigned specific attack objectives and plan their actions while continuing to perform routine duties.

To emulate realistic human behavioral diversity, we follow prior research~\cite{besta2025psychologically}, which finds that assigning personality traits to LLM agents produces identifiable and personalized behavioral patterns. Accordingly, each agent in \chimera is parameterized by a lightweight personality profile based on the Myers-Briggs Type Indicator (MBTI)~\cite{myers1962myers}. This profile influences the agent's decision-making tendencies, communication frequency, writing tone, and risk tolerance. Such traits introduce controlled behavioral variability across agents, leading to emergent diversity in communication styles and work patterns that reflect real organizational dynamics.

\subsection{Threat Scenario Simulation}

After the system environment and agent society are initialized, \chimera simulates realistic organizational operations. Each employee agent operates within a well-defined action space, which includes participating in group meetings, planning daily tasks, and executing assigned activities to achieve organizational objectives. To ensure isolation and prevent any impact on the host system, each agent is executed inside an isolated container that provides an individualized runtime environment. Agent communication is restricted to an internal virtual network, while external interactions such as web browsing are emulated through internal mock servers governed by strict whitelist policies. This design guarantees that no real external network entities are contacted during simulation or attack execution, as further discussed in Section~\ref{sec:ethics}.

During each simulated day, benign agents generate daily schedules and perform tasks accordingly. When communication occurs, agents dynamically adjust subsequent tasks based on discussion outcomes. The simulation proceeds until all assigned objectives are completed. For adversarial agents, \chimera embeds malicious actions into their normal routines, allowing threats to blend with legitimate behavior. Each attack scenario is instantiated following real-world TTPs, which are introduced in Section~\ref{sec:attack}. After analyzing the target employee's schedule, the framework determines optimal timings for attack execution. All attack activities are automatically labeled during logging, enabling consistent and fine-grained ground truth for downstream research.

To formalize the simulation, we define a scenario as:
\[
X = (E, R, S, G, T),
\]
where $E$ denotes the set of employees, $R$ denotes the set of roles with mapping $r:E\rightarrow R$, $S$ denotes the available systems, $G$ denotes the organizational goal, and $T$ denotes the simulation duration. At time step $t$, the organizational state is:
\[
s_t = (\textit{artifacts}, \textit{plan\_status}, \textit{sys\_logs}),
\]
where \textit{artifacts} denote the current organizational assets in the simulation, including systems and documents. The \textit{plan\_status} records each agent's task plan and execution state, including completed and pending tasks. The \textit{sys\_logs} store all monitored application-level and system-level events observed up to time~$t$. 
The simulation process evolves three key stages:

\textbf{Plan Generation:} Given a scenario $X$, \chimera generates plans $P$ at multiple temporal resolutions, including monthly, weekly, and daily plans:
 \[
    P=(P_M,P_W,P_D)=\textit{LLM}_{\text{plan}}(X),
    \]
    
Where employees first collaboratively generate a monthly organizational plan $P_M=f_{\text{org}}^{M}(X)$, which is subsequently refined into weekly plans $P_W=f_{\text{org}}^{W}(P_M,X)$. Given the organizational and weekly plans, each employee $e \in E$ derives an individual daily plan $P_D^e=f_{\text{emp}}(e,P_W,P_M,X)$. $P_D=\{P_D^e | e\in E\}$ shows the complete set of daily plans.
   
    % where $P_M$, $P_W$, and $P_D$ denote monthly, weekly, and daily plans.

\textbf{Execution:} At each time step, employee $e$ performs
    \[
    s_{t+1}=\textit{LLM}_{\text{execute}}(p_t^e,s_t,X),
    \]
where $p_t^e$ includes the planned action in $e$'s plan. Each plan encodes a sequence of time-indexed actions together with their semantic descriptions, which determine the intended activity to be executed at the corresponding time step. This execution produces observable behaviors across multiple log modalities.

\textbf{Plan Update:} After each simulated day or communication event, update task schedules and attack strategies to reflect contextual and behavioral changes.

\[
P'=\textit{LLM}_{\text{update}}(P,X,s_t),
\]

\subsection{Agent Memory Management}

\added[id=CR]{To ensure agents within \chimera exhibit context-consistent behaviours and progress steadily toward the organizational goals, we adopt a hybrid memory architecture comprising long-term and short-term components. Specifically, at the end of each simulated day, every agent produces a daily report summarising their task completions, outstanding objectives, and communications. This ``day $t$ summary'' becomes part of the long-term memory and is used as input on day $t+1$ in conjunction with the meeting-derived agenda and schedule update routines, thereby maintaining continuity of strategy and ensuring goal-alignment across multiple simulation days.} 

\added[id=CR]{For short-term memory, \chimera adopts the built-in MAS memory design, which each agent retains the last five interaction turns (including schedule changes, communications, and tool-usage events) in a sliding-window buffer, aligning with memory-management practices such as CAMEL~\cite{li2023camel} and OWL~\cite{owl2025}, which support stateful contextual agents via dedicated memory modules. Note that while the short-term memory performance depends on the underlying foundation model and multi-agent configuration, the daily summarisation mechanism provides a consistent backbone to preserve behavioural coherence over the simulation horizon.}

\subsection{Log Collection}

\begin{figure*}[t]
    \centering
    \includegraphics[width=1\linewidth]{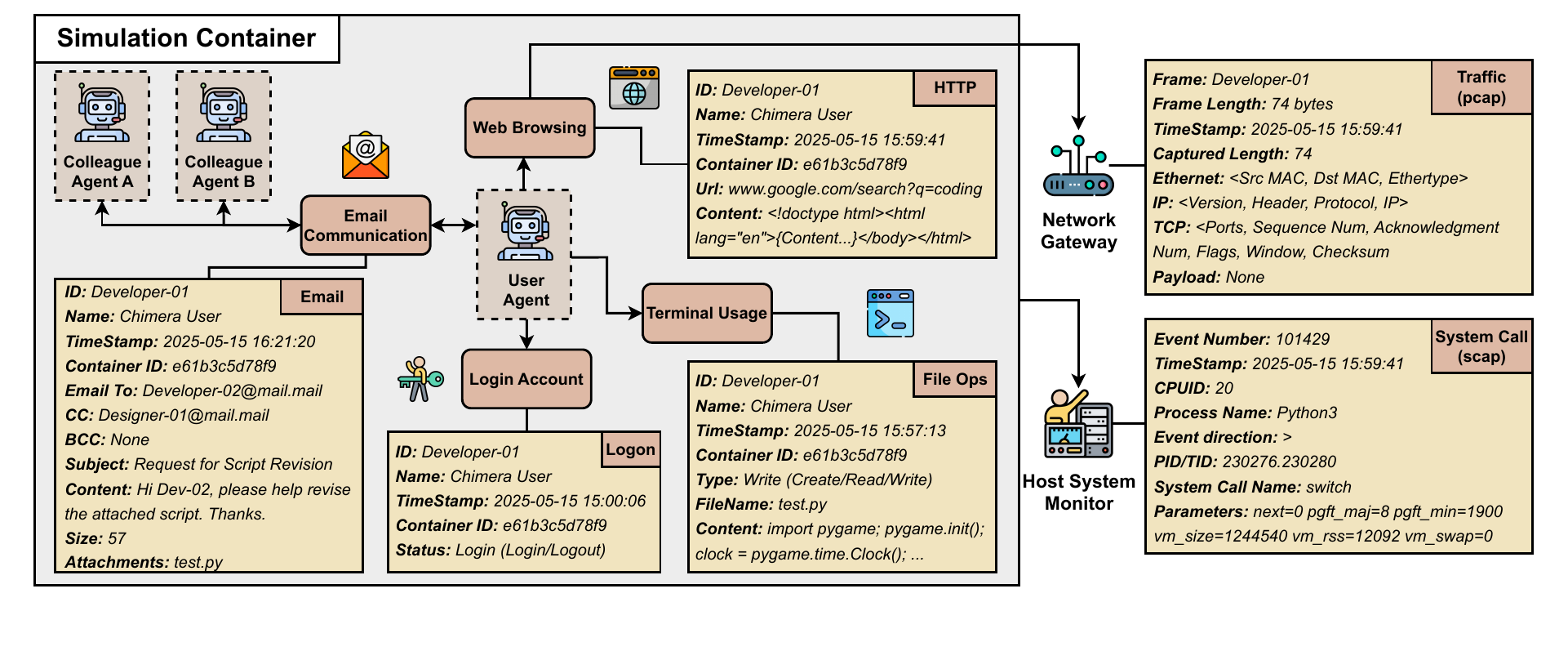}
    \caption{Overview of multi-modal log collection in \dataset.}
    \label{fig:log-format}
\end{figure*}

High-quality logs are essential for developing effective ITD methods. However, achieving unified and consistent logging across heterogeneous system configurations is inherently challenging, as different organizations employ varied software stacks and log formats. The detailed log collection modalities are illustrated in Figure~\ref{fig:log-format}. 

For application-level logs, \chimera aligns its data formats with existing ITD datasets~\cite{kent2015cybersecurity,harilal2017twos}, including email exchanges, login events, file operations, and web browsing activities, to ensure compatibility and adaptability. The collection process is continuously monitored by dedicated application-level collectors, which capture both inter-agent communications and interactions between agents and the environment. These logs record high-level information about each agent, such as user ID, and detailed descriptions of each event.

\added[id=CR]{For system-level logging, \chimera captures the global system behaviors by deploying the complete agent society within an isolated, containerized environment. Network traffic and system call activities are recorded using tcpdump~\cite{pcap} and Sysdig~\cite{scap}, respectively. This setup ensures comprehensive monitoring of all agent activities while maintaining strict separation between simulated entities. Although system-level logs do not explicitly contain user identifiers, the underlying activities triggered by individual employee agents are implicitly reflected in the corresponding system traces, with timestamps mapped to their associated actions.}

Beyond standard log entities, \chimera supports fine-grained tracking of each employee agent’s activities, which are directly linked to scheduled tasks. Crucially, because the logging is integrated into the LLM agents themselves, \chimera records not only observable user-level events but also internal agent operations such as tool invocations, intermediate artifacts (e.g., retrieved webpages, generated code snippets, document edits), and LLM-generated responses. To ensure high temporal resolution, application-level logs are recorded on a simulated daily basis, with every agent action timestamped at sub-second granularity. This precision matches that of prior ITD datasets while offering substantially richer contextual detail.

% Dataset
\section{\dataset Dataset}
\subsection{Dataset Overview}

To assess the effectiveness of \chimera in simulating insider threats under realistic enterprise conditions, we construct a new dataset, \dataset, by deploying the framework across three representative data-sensitive sectors: technology companies, financial corporations, and medical institutions. These domains are selected due to their distinct workflows, access patterns, and insider-risk profiles, thereby covering a broad spectrum of organizational behaviors. 
% \added[id=CR]{A detailed illustration of the simulated organizational scenarios is provided in Appendix~B.}

\subsection{Construction Configuration}

In each scenario, a cohort of 20 employee agents is simulated continuously for one month. This configuration reflects a medium-sized enterprise team commonly observed in prior studies~\cite{homoliak2019insight,randazzo2004insider}. A month-long duration allows the emergence of realistic collaborative workflows, including project milestones, iterative reviews, and temporal variations in activity intensity, while keeping the overall simulation computationally tractable. It is worth noting that \chimera remains fully configurable to support larger scales or longer durations when required. Across the three scenarios, agents collectively perform 15 real-world insider attacks derived from public case reports, interleaved with benign operational activities.

\begin{figure}[t]
    \centering
    \includegraphics[width=1\linewidth]{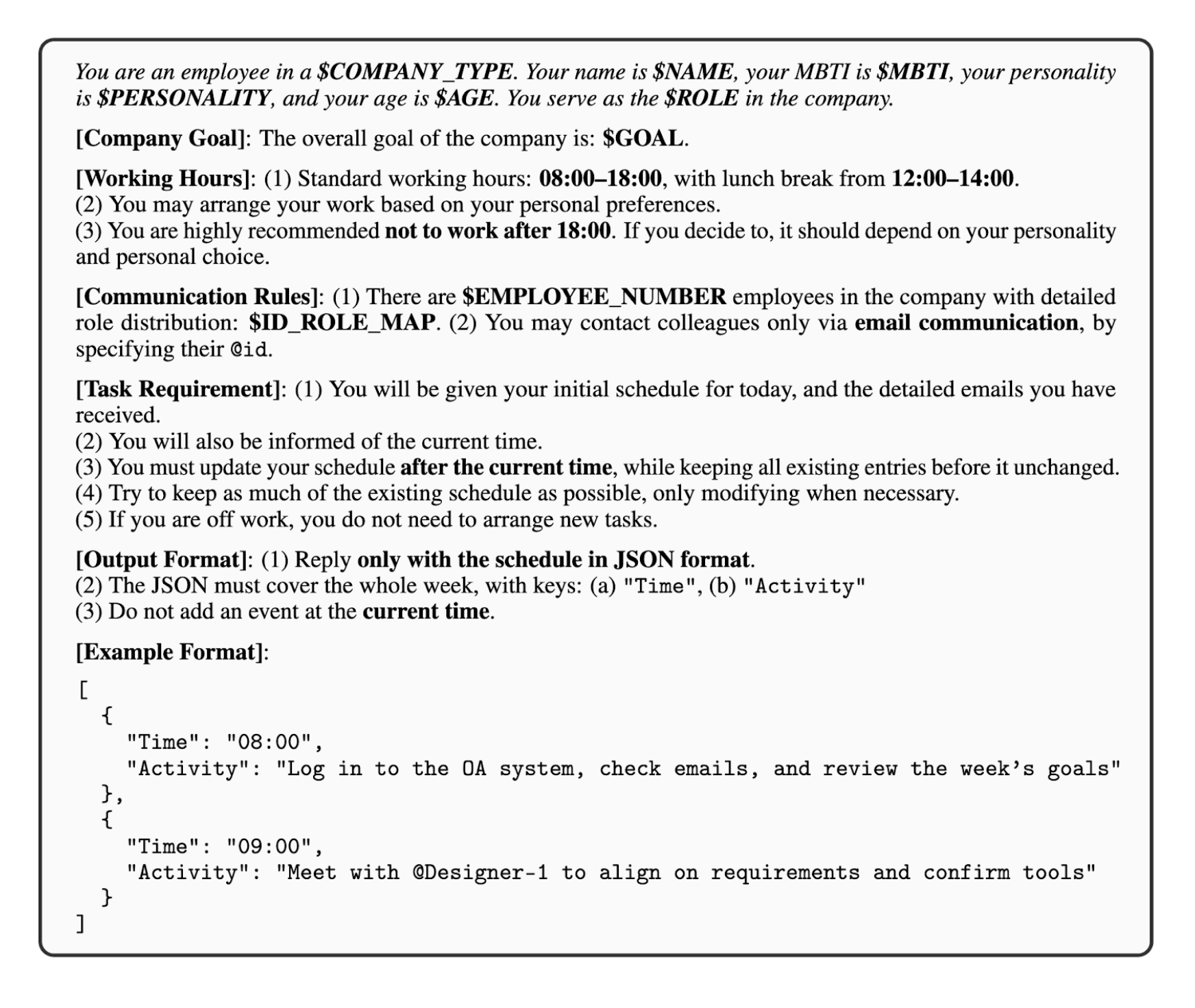}
    \caption{System prompt for agent's daily schedule update.}
    \label{fig:schedule-update}
\end{figure}

\textbf{LLM Configuration.} To ensure consistency, the temperature is fixed at 0 for deterministic operations such as meeting discussions and daily scheduling, while tasks requiring creativity or strategic reasoning (e.g., insider attacks or multi-step planning) use a temperature of 0.7, following the configuration adopted in the CAMEL~\cite{li2023camel} framework. All prompt templates used for dataset construction, including for company profiling, weekly meetings, post-meeting summarization, daily schedule generation and updates, and attacker activity synthesis, are designed following a unified multi-agent coordination paradigm. Inspired by MetaGPT~\cite{hong2023metagpt}, these prompts encode standardized operating procedures (SOPs) as structured prompt sequences to coordinate agents with distinct roles. An example prompt for daily schedule updates is shown in \autoref{fig:schedule-update}.

\textbf{Simulation Scenarios.} To reflect the critical role of ITD in data-sensitive environments, we select three representative data-sensitive organizations for dataset construction.

\begin{itemize}[leftmargin=*]
    \item \textbf{Technology Companies} are particularly susceptible to insider threats due to the rapid pace of innovation and the intensive use of information technology~\cite{wright2025top}. Common insider attack patterns include IP theft~(e.g., stealing source files) and IT sabotage~(e.g., developers spam threat emails). In our simulation, the organizational goal is to develop a game software, i.e., a third-person shooter game.

    \item \textbf{Financial Corporations} are high-value targets due to their management of substantial assets, and insider threats can result in significant financial losses~\cite{randazzo2004insider}. Typical insider scenarios include fraud or data exfiltration, such as an analyst misusing credentials to steal proprietary trading algorithms, or a broker executing unauthorized trades for personal gain. For this scenario, the corporation’s goal is to design a market-neutral statistical arbitrage fund.
    
    \item \textbf{Medical Institutions} manage sensitive patient information, making insider attacks especially damaging~\cite{metomic2025healthcare}. Typical attacks involve illicit access to electronic health records, selling protected health information, or sabotaging medical systems. In our simulation, the institution was tasked with completing electronic health record~(EHR) collection and conducting seasonal influenza trend analysis.
\end{itemize}

% \textbf{Log Modality.} \dataset ensembles with both application and system-level modalities, where login events, emails, web browsing histories, and file operations at the application level~\cite{glasser2013bridging}, and network packets (captured in PCAP format~\cite{packet_capture_appliance}) and system events (in SCAP format~\cite{scap}) at the system level. 

% These parameter settings are chosen to balance realism and reproducibility, reflecting the degree of variability typically observed in human decision-making within enterprise teams.

In total, \dataset comprises approximately 2.0 billion application-level events, including 0.2 billion logons, 0.6 billion email records, 0.8 billion web histories, and 0.4 billion file operations, alongside 4.5 billion network packets and 18.2 billion system log entries, representing over 160 hours of agent activity. Compared with public benchmarks such as \cert and \twos, which contain only coarse-grained application logs, \dataset provides finer temporal granularity, multi-modal coverage, and scale suitable for cross-layer ITD evaluation.

\subsection{Insider Threats}
\label{sec:attack}

Insider threats present a highly heterogeneous class of adversarial activity, given that trusted insiders typically possess a far broader action space than external attackers who exploit vulnerabilities. In our dataset (\dataset), we adopt the unified categorisation framework introduced in prior research \cite{homoliak2019insight}, anchored in the ``who, what, where, when, why and how'' (5W1H) methodology \cite{yang20115w1h}. This approach supports our aim to model as many meaningful insider-attack types as possible and thereby facilitate comprehensive simulation and evaluation.

To this end, we survey public incident databases documenting real-world insider events, including the Data Broker Database \cite{prc2025databroker}, the U.S. Attorney's Office records \cite{usao2025}, and the Federal Bureau of Investigation (FBI) archives \cite{fbi2025}. In total, we incorporate 12 distinct attack types, as well as three hybrid scenarios that combine multiple attack patterns. These hybrid cases were selected on the basis of documented real-world cases from the United States Department of Justice (DOJ).

For each defined attack scenario, we systematically map the corresponding behaviors to the MITRE ATT\&CK for Enterprise framework by assigning relevant TTPs. As shown in \autoref{tab:attack-TTP}, the example IP theft scenario is modeled as a three-stage attack chain. The first stage involves insider recruitment and credential co-option, mapped to Trusted Relationship (T1199) and Valid Accounts (T1078). The second stage captures social engineering-driven privilege acquisition through internal corporate email and improper approval of elevated access, corresponding to Social Engineering (T1566) and Valid Accounts (T1078). The final stage models the exfiltration of source code and design documents to removable media, aligned with Exfiltration over Physical Medium (T1052).

For each stage, we specify a concise procedure, identify observable artifacts such as anomalous inter-peer communications, and enumerate the primary data sources that support detection (e.g., email logs). These tables include Step, Tactic, Technique, Sub technique, Procedure, Observable Evidence, Detection Data Sources, and Impact for all attack scenarios. Such structured TTP mappings enable targeted log collection during attack simulation and establish a clear linkage between simulated attack techniques and the resulting log data for subsequent analysis.

\begin{table*}[t]
\setlength{\abovecaptionskip}{0.05cm}
\setlength{\belowcaptionskip}{0cm}
\centering
\caption{Attacks considered in \dataset. All the attacks are summarized from existing 5W1H taxonomy~\cite{homoliak2019insight} with links to real world cases~\cite{website}.} 
% \fishfix{Add Attack Index matching with the collection result.}}
\label{tab:attacks}
\resizebox{1\linewidth}{!}{
\begin{tabular}{c|c|c|c|c|c}
\hline
\textbf{Attacker} & \textbf{Role}    & \textbf{Goal}                      & \textbf{Target}  & \textbf{Frequency} & \textbf{Purpose} \\ \hline
Traitor                & internal          & IP theft     & OS, Network, App & recurrent          & financial                    \\ \hline
Traitor                & internal          & IP theft     & OS, Network, App & single             & financial                    \\ \hline
Traitor                & internal          & IP theft     & App              & single             & financial                    \\ \hline
Traitor                & internal/external & sabotage                           & App              & single             & financial/personal           \\ \hline
Traitor                & internal/external & sabotage                           & OS               & single             & financial/personal           \\ \hline
Traitor                & internal/external & sabotage                           & OS, Network               & single             & financial/personal           \\ \hline
Masqueraders           & internal/external & fraud                              & App              & single             & financial                    \\ \hline
Masqueraders           & internal/external & fraud                              & OS             & recurrent          & financial/personal            \\ \hline
Masqueraders           & internal/external & IP theft     & OS, Network      & recurrent          & financial/political          \\ \hline
Masqueraders           & internal/external & IP theft     & OS, Network, App & recurrent          & financial                    \\ \hline
Unintentional User       & internal          & data leak                          & OS, Network      & single             & personal                     \\ \hline
Unintentional User       & internal          & IP theft     & App              & recurrent          & personal                     \\ \hline
Miscellaneous          & internal          & data exfiltration                  & App              & recurrent          & financial                    \\ \hline
Miscellaneous          & internal          & data exfiltration                  & OS, Network      & recurrent          & financial                    \\ \hline
Miscellaneous          & internal/external & data exfiltration, system takeover & OS, Network, App & recurrent          & political                    \\ \hline
\end{tabular}
}
\end{table*}

\begin{table*}[t]
\setlength{\abovecaptionskip}{0.05cm}
\setlength{\belowcaptionskip}{0cm}
\centering
\caption{Example ATT\&CK TTPs mapping of the insider IP theft attack.} 
% \fishfix{Add Attack Index matching with the collection result.}}
\label{tab:attack-TTP}
\resizebox{\linewidth}{!}{
\begin{tabular}{|c|c|c|c|c|c|}
\hline
\textbf{Step} & \textbf{Tactic}      & \textbf{Technique} & \textbf{Sub-technique} & \textbf{Procedure}                                                                                                    & \textbf{Detection Data Sources}            \\ \hline
1             & Initial Access       & T1199, T1078       & -                      & Insider recruits colleagues to obtain access or credentials                                                           & Emails, System Logs                        \\ \hline
2             & Privilege Escalation & T1566, T1078       & -                      & Perpetrator uses email to request resource administrators with admin-level access          & Emails, Logon                              \\ \hline
3             & Exfiltration         & T1052              & T1052.001              & Download sensitive files, transfer to private cloud, remove from premises & System Logs, Traffic, File Operation, HTTP \\ \hline
\end{tabular}
}
\end{table*}

% Experiment
\section{Evaluation}

\subsection{Overview}

Based on \dataset, we conduct a comprehensive evaluation to assess the dataset quality and the effectiveness of existing ITD methods in detecting insider threats in \dataset. Our evaluation consists of the following steps:

\begin{itemize}[leftmargin=*]
    \item \textbf{Quality Evaluation.} 
    We conduct a human study and quantitative analysis to compare \dataset against existing datasets (i.e., \cert and \twos) in terms of realism.     
    \item \textbf{ITD Evaluation.} We benchmark four representative machine learning based ITD methods~(SVM, CNN, GCN, and DS-IID) using \dataset. We further evaluate the cross-dataset generalization~(e.g., training on the Chimera-Tech vs. testing on Chimera-Finance or \cert) to investigate how well models trained on one data distribution can detect threats in another.

\end{itemize}

\subsection{Evaluation Setup}

\textbf{Dataset.} Our evaluation covers key datasets including our constructed \dataset, \cert insider threat dataset~(v6.2), and \twos. For human study, we follow previous research~\cite{yang2022natural} and randomly sample 100 log entries from each dataset as the study subjects. For the ITD evaluation, for each dataset, we separate it into training, validation, and test sets. To ensure fair evaluation and prevent overfitting, we set aside 10\% of each dataset as a test set~(never seen during training). The remaining data is split into training~(80\%) and validation~(20\%) sets. We maintain the original class proportions~(normal vs. malicious instances) in all splits. For cross-dataset experiments, we similarly use 10\% of the target dataset as test data and train on the entirety of the source dataset’s training split. All datasets are preprocessed into a common feature format, allowing the baseline models to be applied uniformly. In particular, we extract user-day behavioral features following prior work~\cite{homoliak2019insight} for \cert and \dataset, which includes aggregating log characteristics for all the log events per user per day. 
% For \twos, as the dataset has a different format, we only use it in the human study.

\textbf{ITD Models.} We evaluate four established ITD methods. SVM~\cite{janjua2020handling} is a kernel-based binary classifier that separates normal and malicious user day profiles using a radial basis function kernel. Temporal CNN~\cite{hong2022graph} employs a convolutional architecture to process fixed-length sequences of daily user behavior vectors, using stacked convolutional layers with ReLU activation and max pooling to extract temporal patterns, followed by fully connected classification layers. GCN~\cite{hong2022graph} applies graph convolution to propagate node features across neighborhoods and performs node-level classification. As a representative deep learning approach, we include Deep Synthesis Insider Intrusion Detection (DS-IID)~\cite{kotb2025novel}, which combines an LSTM-based model with an autoencoder to jointly learn normal log reconstruction and attack classification. We use the publicly available implementation of DS-IID and follow the original paper for model configuration and hyperparameter settings.

\textbf{Foundation LLMs and Agentic Frameworks.} \chimera relies on LLMs as cognitive engines to simulate human-like organizational behaviors and agentic operations. Importantly, \chimera is designed as a flexible multi-agent paradigm for insider threat simulation. This flexibility allows it to be adapted to multiple LLM agent frameworks and foundation models. In our experiments, we employ three models, including Google Gemini‑2.0‑Flash, OpenAI GPT‑4o, and DeepSeek V3.

We implement \chimera based on famous multi-agent frameworks CAMEL~\cite{li2023camel} and OWL~\cite{owl2025}. The multi-agent frameworks provide a role-playing communicative architecture that supports multiple agents with individual roles and coordinated interactions. Since the design of \chimera is based on these standardized multi-agent platforms, migration to other agent frameworks such as AutoGen~\cite{wu2024autogen} or custom MCP‑enabled systems requires only minimal engineering effort.

\subsection{Evaluation Metrics}

For the ITD baseline evaluation, to mitigate the effect of randomness introduced by the model training process and improve the reliability of the results, we repeat all experiments five times and report the average results. We evaluate the performance of ITD methods using three standard metrics: \textit{Precision}, \textit{Recall}, and \textit{F1-Score}. Specifically, \textit{Precision} denotes the proportion of true attacks among all instances classified as attacks. \textit{Recall} measures the proportion of actual attacks that are correctly identified by the model. The \textit{F1-Score} is the harmonic mean of \textit{Precision} and \textit{Recall}, providing a balanced measure of the classifier’s effectiveness. A higher F1-score indicates stronger overall performance. We define a True Positive~(TP) as an attack log that is correctly labeled as an attack, while a FP refers to a normal log that is incorrectly labeled as an attack. The formulas for each metric are summarized as follows:
\begin{align}
\text{Precision}  &= \frac{TP}{TP + FP}\\
\text{Recall}     &= \frac{TP}{TP + FN}\\
\text{F1-Score}   &= \frac{2 \times \text{Precision} \times \text{Recall}} {\text{Precision} + \text{Recall}}
\end{align}

% Experiment Result
\section{Evaluation Results}

\subsection{Quality of \dataset Dataset}

% \added[id=CR]{Given the data-centric nature of ITD tasks, to ensure that \dataset realistically reflects real-world enterprise environments, we design our validation procedure to combine expert assessment with quantitative realism analysis. Our design refers to prior work in synthetic log generation and realism evaluation~\cite{landauer2020have,graham2018overview}. Specifically, we first invited domain experts with professional experience in corporate operations to qualitatively assess the realism and practicality of sampled logs, which ensures the evaluation aligns with actual organizational workflows rather than purely synthetic behaviours. In addition, \fishfix{we complement expert impressions with quantitative analyses of behavioural distributions, including temporal activity periodicity, inter-modality correlations (e.g., between email, login, and file events), and entropy of user actions metrics} that mirror human operational patterns and cross-channel dependencies observed in real enterprises.}

\added[id=CR]{Given the data-centric nature of ITD tasks, we design the validation of \dataset to jointly incorporate expert assessment and quantitative realism analysis, ensuring that it faithfully reflects real-world enterprise behaviors. Following prior work on synthetic log realism evaluation~\cite{landauer2020have,graham2018overview}, we first invited domain experts with experience in corporate security operations to qualitatively assess sampled logs for their realism and practicality, ensuring alignment with genuine organizational workflows rather than purely synthetic behavior. Beyond expert impressions, we conduct quantitative analyses across multiple datasets. Specifically, we compared \textit{event count distributions} by constructing normalized hourly activity histograms, \textit{behavioral entropy} to measure user action diversity, and \textit{sequence complexity} as an estimate of behavioral predictability. Given the differing log formats across datasets, we harmonize comparable event types of existing datasets to ensure fair cross-domain comparison. These metrics collectively capture temporal regularity, cross-modal coordination, and human-like variability, providing both qualitative and quantitative evidence of \dataset's realism.}

\textbf{Human Study.} We aim to evaluate the realism and practical utility of our collected dataset, specifically examining whether the logs accurately reflect real-world activities and how likely these scenarios are to occur in actual enterprise environments. Given the large volume of log entries in \dataset, it is infeasible to manually inspect every entry. To address this, we follow established research methodologies~\cite{yang2022natural,neyman1992two} and apply stratified sampling. Specifically, we select 100 log entries from each dataset, including \dataset, \cert, and \twos, for human assessment.

Our sampling strategy adopts stratification along three dimensions (i.e., dataset, log modality, and behavior). For each dataset, we maintain a fixed total of 100 sampled entries. To maximize interpretability for human experts, we restrict our analysis to four application-level log modalities (i.e., logon, email, web history, and file operations), while excluding system call and network flow records, as these are not human-readable and would hinder consistent evaluation within practical time constraints. Within each dataset, we further balance the sample by selecting an equal number of benign and attack entries, resulting in 50 benign and 50 attack logs per dataset. We employ a stratified sampling strategy inspired by Neyman allocation~\cite{neyman1992two}. For each behavior class, the selected log entries are distributed across modalities in proportion to their relative contribution. Since within-modality variance is unavailable, we approximate Neyman allocation using proportional allocation based on modality size and round the resulting counts to the nearest integer.

% \[
% n_{select}^{(c)}
% \;=\;
% n^{(c)} \cdot
% \frac{N_{m}^{(c)}\, S_{m}^{(c)}}
% {\displaystyle \sum_{j \in \mathcal{M}} N_{j}^{(c)}\, S_{j}^{(c)}},\ m \in \mathcal{M},\ c \in \{\mathrm{B,A}\}.
% \]

% We adopt a stratified sampling strategy inspired by Neyman allocation~\cite{neyman1992two}. For each class, the number of selected log entries is allocated across modalities in proportion to their relative contribution. In our study, we sample a total of 50 entries per class. Since within-modality variance is unavailable, we approximate Neyman allocation using proportional allocation based on modality size and round the selected counts to the nearest integer.

% Sample allocation follows Neyman allocation~\cite{neyman1992two} principles, whereby the number of selected entries for each stratum is determined based on within-stratum variability. Specifically, $M$ denotes the total number of log modalities, $N_m^{(c)}$ represents the total number of logs in modality $m$ for class $c$, and $n^{(c)}$ is the target sample size for each class $c \in \{\mathrm{Benign, Attack}\}$, which in our study is 50. Since the within-stratum standard deviation $S_m^{(c)}$ is not available a priori, we revert to proportional allocation based on stratum size, rounding $n_{select}^{(c)}$ to the nearest integer for each modality and class.

We invited five independent experts, each with at least five years of experience in security and artificial intelligence, from esteemed universities and leading security companies. All of the experts possess deep familiarity with insider threat scenarios within large corporations. Each expert individually evaluated the 100 sampled log snippets for each dataset. The experts were presented with the same set of log entries, which were shuffled to prevent bias. For each log entry, the experts rated its realism and practicality using a five-point Likert scale, drawing on their professional expertise. The evaluation questions include \textit{``The timestamps and event frequency align with a typical workday rhythm''} and \textit{``Overall, I would be inclined to believe that this log segment was captured in a real production environment.''} A rating of 1 indicated that the log content was not realistic, while a rating of 5 indicated strong agreement that the log patterns closely reflected real-world activities.

% ~\footnote{The comprehensive set of evaluation questions is available on our anonymous project website~\cite{website}.}

\begin{figure}[t]
    \centering
    \includegraphics[width=.8\linewidth]{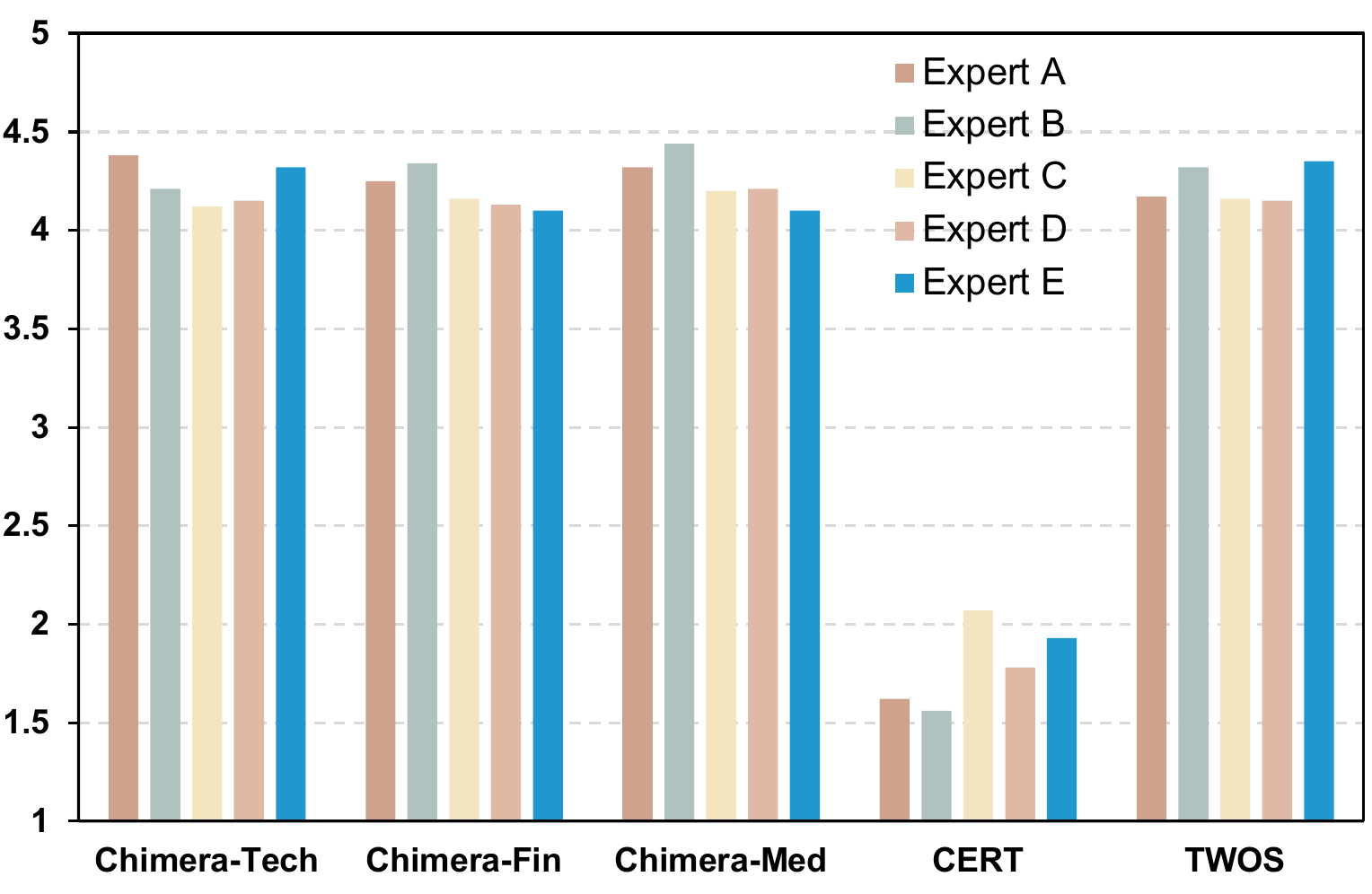}
    \caption{Realism study result by human experts. The y-axis corresponds to the average ratings (1 refers to very unrealistic; 5 refers to very realistic).}
    \label{fig:humanstudy}
\end{figure}

We calculate the average ratings given to each of the questions in the dataset per participant, and present the results in Figure~\ref{fig:humanstudy}. The x-axis distinguishes each dataset we used for evaluation, where we separate the three scenarios of \dataset into three rating candidates, and the y-axis shows the average ratings. The results demonstrate that all three organizational scenarios simulated in \dataset received expert recognition for their high degree of realism, comparable to the real-world \twos dataset. Specifically, the five participating experts awarded an average realism score of 4.20 to \dataset, which is only marginally lower than the 4.25 average score assigned to \twos. This suggests that experts perceive the logs in both datasets as highly natural and realistic. In contrast, the \cert dataset received consistently low scores, with an average of 1.78, reflecting experts' views that its logs lack realism. The primary criticism was that \cert focuses primarily on system graph construction and populates logs with randomly generated, semantically impoverished content. We quantify inter-rater agreement on the 5-point realism scale using Krippendorff’s alpha~\cite{krippendorff2004reliability} ($\alpha$) to demonstrate the consistency and significance of expert ratings. We report both the point estimate and bootstrap 95\% confidence intervals for $\alpha$. Our analysis indicates high inter-rater reliability, with $\alpha = 0.87 $, and the corresponding confidence interval supports the robustness of these results. This high level of agreement demonstrates that expert ratings are highly consistent, thereby confirming the reliability of the evaluation.

\begin{figure*}[t]
    \centering
    \includegraphics[width=1\linewidth]{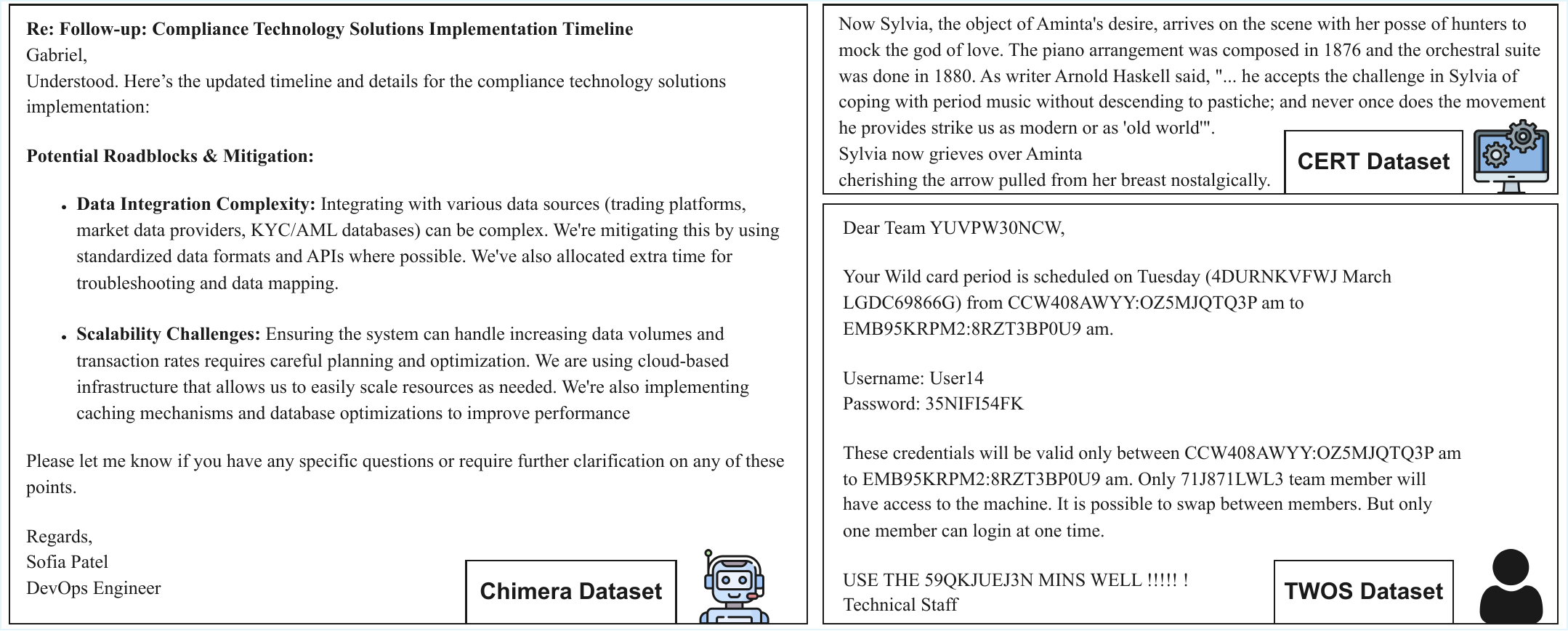}
    \caption{Example of the email communication data in three datasets.}
    \label{fig:email}
\end{figure*}

As illustrated by the example email in Figure~\ref{fig:email}, the email content generated by \chimera is both comprehensive and logically coherent, clearly surpassing \twos and \cert in semantic richness and linguistic naturalness. Specifically, the example of \chimera depicts a realistic corporate communication: the sender references concrete technical components, outlines mitigation strategies, invites further discussion, and closes with a professional signature.

In contrast, the \cert dataset primarily contains syntactically valid but semantically shallow text that lacks contextual continuity. The \twos dataset, although grounded in genuine human interactions, is subject to extensive redaction of sensitive information, which results in brief and fragmented exchanges. These comparisons indicate that \chimera not only generates plausible individual messages but also preserves coherent organizational workflows (e.g., meeting planning, follow-up communication). This example illustrates how \chimera bridges the gap between abstract behavioral modeling and realistic enterprise activity.

\begin{figure}[t]
    \setlength{\abovecaptionskip}{0.05cm}
    \setlength{\belowcaptionskip}{0cm}
    \centering
    \begin{subfigure}{.8\linewidth}
        \includegraphics[width=\linewidth]{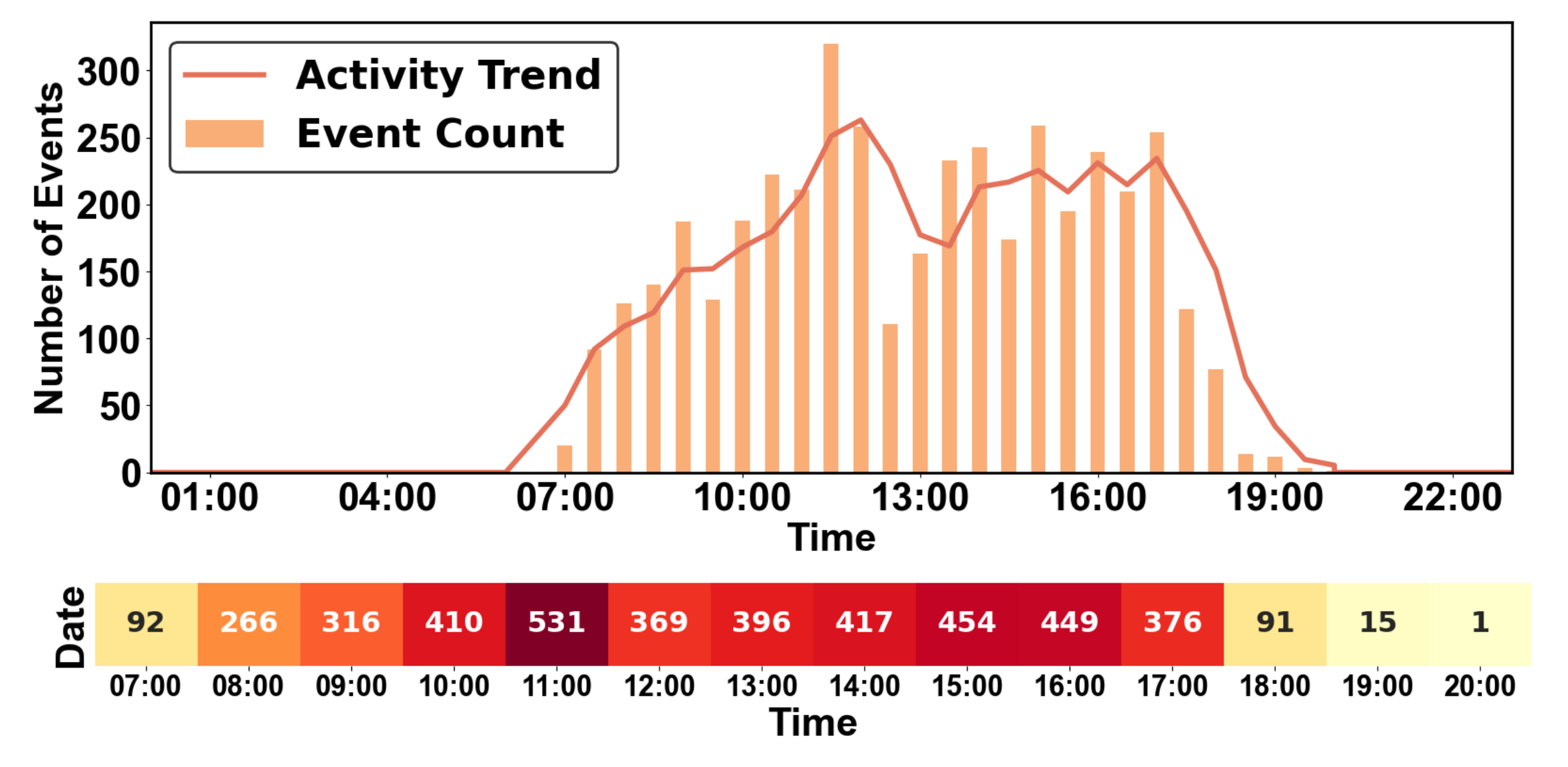}
        \caption{Daily benign employee activity of \dataset.}
        \label{fig:activity-chimera}
    \end{subfigure}
    \begin{subfigure}{.8\linewidth}
        \includegraphics[width=\linewidth]{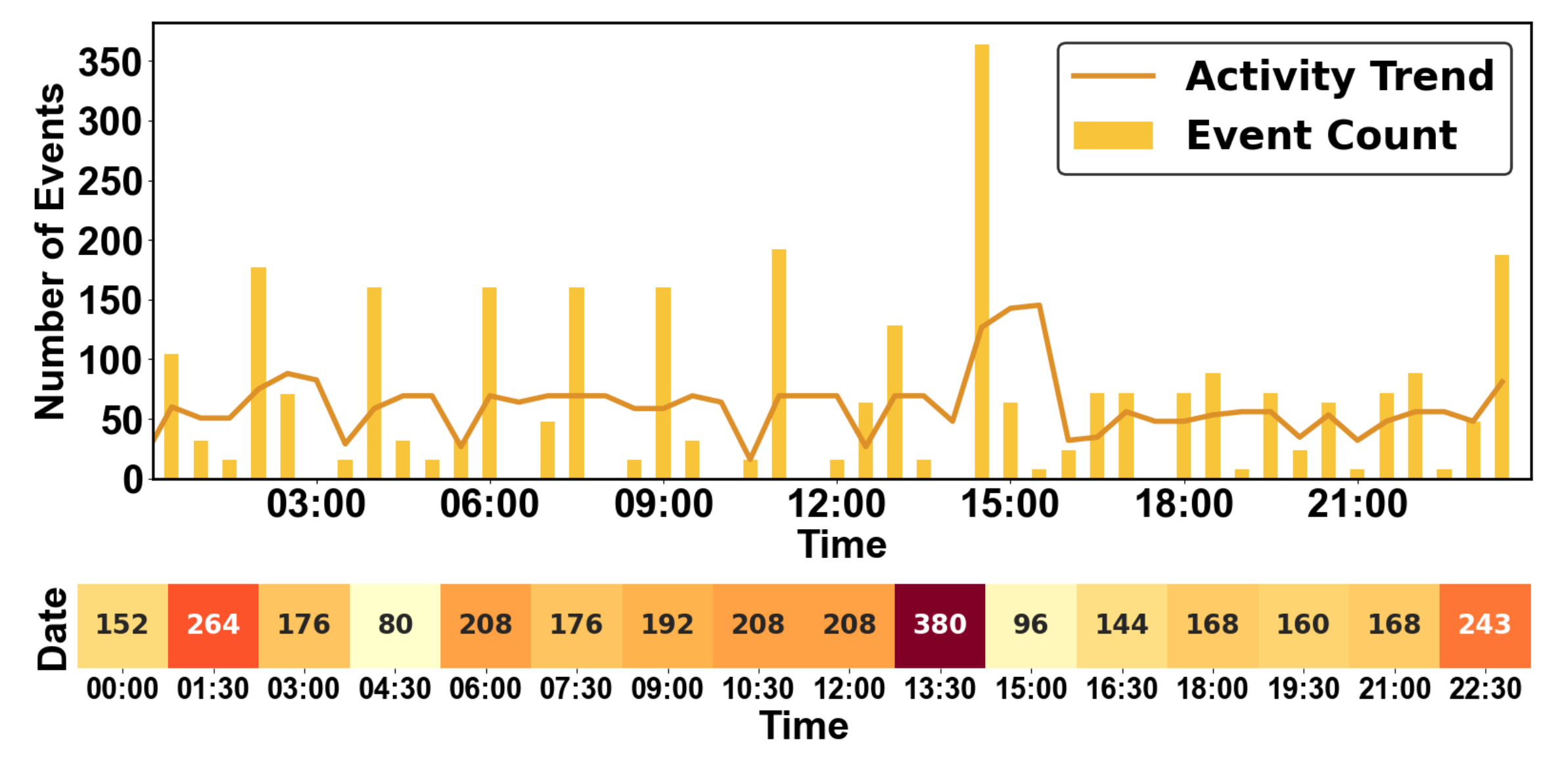}
        \caption{Daily benign employee activity of \twos.}
        \label{fig:activity-twos}
    \end{subfigure}
    \begin{subfigure}{.8\linewidth}
        \includegraphics[width=\linewidth]{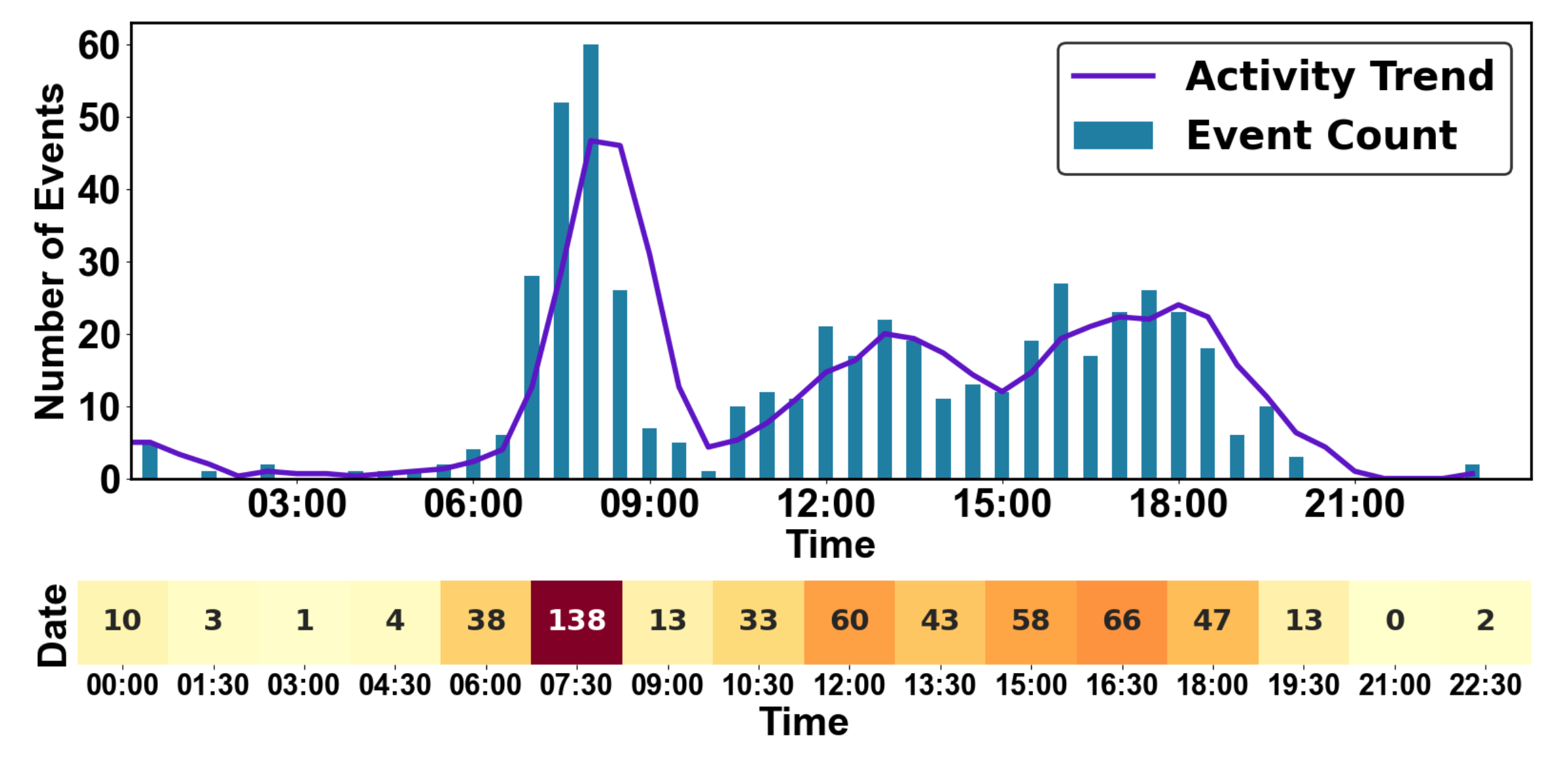}
        \caption{Daily benign employee activity of \cert.}
        \label{fig:activity-cert}
    \end{subfigure}
    \caption{Comparison of benign employee activities among different ITD datasets.}
    \label{fig:activity-comparison}
\end{figure}

\textbf{Quantitative Analysis.} 
We further conduct quantitative analyses across five datasets (\dataset, \cert, \twos, \textit{LANL}, and \textit{OpTC}) to assess the realism of benign activities and the behavioral diversity relevant to ITD. For each dataset, we sample simulated daily activities and construct normalized hourly activity histograms to capture the temporal distribution of user events over a 24-hour period. As shown in \autoref{fig:activity-comparison}, \dataset exhibits clear working-hour periodicity, including pronounced morning start-up peaks, reduced activity during lunch hours, and gradual declines in the evening. These patterns indicate realistic and consistent temporal regularity.

We then quantify behavioral entropy to measure the diversity of user actions under the same log modality. 
For each user $u$, entropy is computed as 
\[
H_u = -\sum_{a \in A} p_u(a)\log p_u(a),
\]
where $A$ denotes the action vocabulary (e.g., logon, email) and $p_u(a)$ the empirical probability of action $a$ by user $u$. 
We compare the logs of \chimera and \cert to ensure consistency across modalities. Our analysis shows that \dataset consistently exhibits higher entropy, indicating richer and more diverse user behaviors than \cert, which averages 2.12 bits. We also compute the inter-event correlation across different modalities to evaluate the degree of cross-modal coherence. The results show that Chimera maintains a high rate of co-occurrence among modalities, with an average cross-modal probability of 0.66, suggesting strong behavioral consistency across different log types.

Finally, we evaluate {sequence complexity} as an indicator of behavioral regularity and compressibility. Following prior studies on log compression and sequence complexity~\cite{li2024logshrink,wei2025logcrisp}, we approximate the Kolmogorov complexity of each user's event stream with the normalized compression ratio:
\[
C_u = 1 - \frac{L_\text{compressed}(S_u)}{L_\text{raw}(S_u)},
\]
where $S_u$ denotes the serialized log sequence for user $u$ and $L_\text{compressed}$ is obtained using gzip compression. Higher $C_u$ values indicate lower redundancy and greater behavioral diversity. Our results show that \dataset achieves higher complexity (77.9\%) than other datasets (\cert 41.2\%, \twos 21.6\%), reflecting non-repetitive activity sequences.

\begin{tcolorbox}[size=title,opacityfill=0.1,breakable]

\textbf{Finding 1:} \dataset demonstrates similar authenticity comparable to the real-world \twos dataset, which is a significant improvement over \cert, whose logs lack meaningful semantic content. At the same time, \dataset retains the advantages of \cert in terms of realistic activity patterns, which are not present in \twos. This combination highlights the practical potential of \dataset for use in the ITD domain.
\end{tcolorbox}

\subsection{Effectiveness of Existing ITD in \dataset}

\begin{table*}[t]
\centering
\caption{Evaluation results of ITD models across different scenarios in the \chimera and \cert datasets. Best and worst results for each dataset are highlighted in green and pink, respectively.}
\label{tab:rq2}
\resizebox{\linewidth}{!}{
\begin{tabular}{c|cccc|cccc|cccc|cccc}
\hline
\multirow{2}{*}{\diagbox{\textbf{Baseline}}{\textbf{Dataset}}} & \multicolumn{4}{c|}{\textbf{Chimera-Tech}}  & \multicolumn{4}{c|}{\textbf{Chimera-Finance}}     & \multicolumn{4}{c|}{\textbf{Chimera-Medical}}     & \multicolumn{4}{c}{\textbf{\cert}}                    \\ \cline{2-17} 
                        & Acc & \multicolumn{1}{l}{Pre} & Recall & F1 & Acc & \multicolumn{1}{l}{Pre} & Recall & F1 & Acc & \multicolumn{1}{l}{Pre} & Recall & F1 & Acc & \multicolumn{1}{l}{Pre} & Recall & F1 \\ \hline
\textbf{SVM}                     & 0.751  & 0.679                    & 0.823   & 0.744 & 0.749 & 0.753                   & 0.639  & \worst 0.691 & 0.755 & 0.743                   & 0.692  & 0.717 & 0.873 & 0.884                   & 0.931  & \worst 0.907 \\ \hline
\textbf{CNN}                     & 0.864  & 0.890                    & 0.739   & 0.808 & 0.794 & 0.740                   & 0.891  & \best 0.809 & 0.851 & 0.858                   & 0.780  & 0.817 & 0.923 & 0.891                   & 0.959  & 0.924 \\ \hline
\textbf{GCN}                     & 0.697  & 0.674                    & 0.727   & \worst 0.699 & 0.755 & 0.669                   & 0.749  & 0.707 & 0.669 & 0.671                   & 0.736  & \worst 0.702 & 0.913 & 0.927                   & 0.943  & 0.935 \\ \hline
\textbf{DS-IID}                  & 0.826  & 0.727                    & 0.949   & \best 0.823 & 0.783 & 0.781                   & 0.792  & 0.786 & 0.904 & 0.857                   & 0.784  & \best 0.819 & 0.971 & 0.960                   & 0.950  & \best 0.955 \\ \hline
\end{tabular}
}
\end{table*}

We evaluate the effectiveness of existing ITD methods on our \chimera-generated dataset. The average performance of each method is summarized in Table \ref{tab:rq2}. Overall, existing ITD methods can identify most insider threats, but their performance varies substantially. While they achieve strong results on the \cert dataset, their performance fluctuates by as much as 20\% on \dataset, highlighting the increased complexity and challenge presented by our dataset compared to the purely simulated \cert logs. Notably, Chimera-Finance emerges as the most challenging scenario, suggesting that internal threat behaviors within financial institutions are more deeply concealed than in other contexts. This finding underscores the urgent need for the development of more robust ITD methods tailored to real-world environments. Furthermore, we evaluate the generalization capability of existing ITD methods under distributional shifts. Specifically, we train each ITD model on one dataset (e.g., Chimera-Tech), and evaluate it on a different dataset (e.g., \cert).

As shown in \autoref{tab:rq3}, distributional shifts significantly degrade the performance of existing ITD models. For example, the F1 score of DS-IID drops by 49.1\% when evaluated out of distribution on Chimera-Finance compared to in-distribution performance on \cert, highlighting the need for ITD methods that are robust to distributional shifts. Moreover, models trained on \cert show poor generalization, often yielding FP rates approaching 100\% when deployed in other environments, which exposes the limitations of ITD systems trained solely on synthetic datasets.

\begin{tcolorbox}[size=title,opacityfill=0.1,breakable]

\textbf{Finding 2:} \dataset is more challenging for ITD models compared with \cert, and models trained on the \dataset demonstrate better generalization. While ITD methods are effective in detecting insider threats, all methods experience significant performance declines when faced with distribution shifts.
\end{tcolorbox}

\begin{table*}[t]
\centering
\caption{Distribution shift evaluation of ITD models on different datasets.}
\label{tab:rq3}
\resizebox{\linewidth}{!}{
\begin{tabular}{c|c|cccc|cccc|cccc|cccc}
\hline
\multirow{2}{*}{\textbf{Train Dataset}}   & \multirow{2}{*}{\textbf{Test Dataset}} & \multicolumn{4}{c|}{\textbf{SVM}} & \multicolumn{4}{c|}{\textbf{CNN}} & \multicolumn{4}{c|}{\textbf{GCN}} & \multicolumn{4}{c}{\textbf{DS-IID}} \\ \cline{3-18} 
                                          &                                        & Acc    & Pre    & Recall  & F1    & Acc    & Pre    & Recall  & F1    & Acc    & Pre    & Recall  & F1    & Acc     & Pre    & Recall  & F1     \\ \hline
\multicolumn{1}{c|}{\multirow{3}{*}{\textbf{Chimera-Tech}}}    & Chimera-Finance & 0.700  & 0.550  & 0.600   & $0.574^{\textcolor{red}{\downarrow 0.170}}$ & 0.700  & 0.550  & 0.600   & $0.574^{\textcolor{red}{\downarrow 0.234}}$ & 0.755  & 0.627  & 0.772   & $0.692^{\textcolor{red}{\downarrow 0.008}}$ & 0.821   & 0.850  & 0.757   & $0.801^{\textcolor{red}{\downarrow 0.023}}$  \\
\multicolumn{1}{c|}{}                                          & Chimera-Medical & 0.688  & 0.500  & 0.500   & $0.500^{\textcolor{red}{\downarrow 0.244}}$ & 0.688  & 0.500  & 0.500   & $0.500^{\textcolor{red}{\downarrow 0.308}}$ & 0.771  & 0.251  & 0.609   & $0.356^{\textcolor{red}{\downarrow 0.344}}$ & 0.816   & 0.851  & 0.752   & $0.798^{\textcolor{red}{\downarrow 0.025}}$  \\
\multicolumn{1}{c|}{}                                          & CERT            & 0.354  & 0.880  & 0.500   & $0.638^{\textcolor{red}{\downarrow 0.106}}$ & 0.357  & 0.926  & 0.317   & $0.472^{\textcolor{red}{\downarrow 0.335}}$ & 0.298  & 0.617  & 0.251   & $0.357^{\textcolor{red}{\downarrow 0.343}}$ & 0.811   & 0.850  & 0.761   & $0.803^{\textcolor{red}{\downarrow 0.020}}$  \\ \hline
\multicolumn{1}{c|}{\multirow{3}{*}{\textbf{Chimera-Finance}}} & Chimera-Tech    & 0.700  & 0.650  & 0.700   & $0.674^{\textcolor{red}{\downarrow 0.017}}$ & 0.700  & 0.650  & 0.800   & $0.717^{\textcolor{red}{\downarrow 0.091}}$ & 0.699  & 0.457  & 0.531   & $0.491^{\textcolor{red}{\downarrow 0.216}}$ & 0.822   & 0.850  & 0.658   & $0.742^{\textcolor{red}{\downarrow 0.045}}$  \\
\multicolumn{1}{c|}{}                                          & Chimera-Medical & 0.667  & 0.500  & 0.500   & $0.500^{\textcolor{red}{\downarrow 0.191}}$ & 0.667  & 0.500  & 0.667   & $0.571^{\textcolor{red}{\downarrow 0.237}}$ & 0.880  & 0.698  & 0.705   & $0.702^{\textcolor{red}{\downarrow 0.006}}$ & 0.831   & 0.850  & 0.768   & $0.776^{\textcolor{red}{\downarrow 0.010}}$  \\
\multicolumn{1}{c|}{}                                          & CERT            & 0.388  & 0.933  & 0.357   & $0.516^{\textcolor{red}{\downarrow 0.175}}$ & 0.386  & 0.933  & 0.355   & $0.515^{\textcolor{red}{\downarrow 0.294}}$ & 0.340  & 0.693  & 0.302   & $0.421^{\textcolor{red}{\downarrow 0.286}}$ & 0.804   & 0.850  & 0.653   & $0.739^{\textcolor{red}{\downarrow 0.048}}$  \\ \hline
\multicolumn{1}{c|}{\multirow{3}{*}{\textbf{Chimera-Medical}}} & Chimera-Tech    & 0.500  & 0.250  & 0.250   & $0.250^{\textcolor{red}{\downarrow 0.467}}$ & 0.563  & 0.250  & 0.250   & $0.250^{\textcolor{red}{\downarrow 0.567}}$ & 0.820  & 0.667  & 0.704   & $0.685^{\textcolor{red}{\downarrow 0.017}}$ & 0.827   & 0.850  & 0.763   & $0.804^{\textcolor{red}{\downarrow 0.015}}$  \\
\multicolumn{1}{c|}{}                                          & Chimera-Finance & 0.583  & 0.250  & 0.330   & $0.284^{\textcolor{red}{\downarrow 0.432}}$ & 0.583  & 0.250  & 0.333   & $0.286^{\textcolor{red}{\downarrow 0.531}}$ & 0.859  & 0.678  & 0.735   & $0.701^{\textcolor{red}{\downarrow 0.001}}$ & 0.817   & 0.850  & 0.763   & $0.804^{\textcolor{red}{\downarrow 0.015}}$  \\
\multicolumn{1}{c|}{}                                          & CERT            & 0.303  & 0.944  & 0.243   & $0.387^{\textcolor{red}{\downarrow 0.330}}$ & 0.293  & 0.950  & 0.229   & $0.370^{\textcolor{red}{\downarrow 0.448}}$ & 0.264  & 0.724  & 0.204   & $0.319^{\textcolor{red}{\downarrow 0.383}}$ & 0.813   & 0.851  & 0.652   & $0.738^{\textcolor{red}{\downarrow 0.081}}$  \\ \hline
\multicolumn{1}{c|}{\multirow{3}{*}{\textbf{CERT}}}            & Chimera-Tech    & 0.300  & 0.300  & 1.000   & $0.462^{\textcolor{red}{\downarrow 0.445}}$ & 0.300  & 0.300  & 1.000   & $0.462^{\textcolor{red}{\downarrow 0.462}}$ & 0.300  & 0.300  & 1.000   & $0.462^{\textcolor{red}{\downarrow 0.473}}$ & 0.341   & 0.354  & 0.705   & $0.471^{\textcolor{red}{\downarrow 0.484}}$  \\
\multicolumn{1}{c|}{}                                          & Chimera-Finance & 0.300  & 0.300  & 1.000   & $0.462^{\textcolor{red}{\downarrow 0.445}}$ & 0.300  & 0.300  & 1.000   & $0.462^{\textcolor{red}{\downarrow 0.462}}$ & 0.300  & 0.300  & 1.000   & $0.462^{\textcolor{red}{\downarrow 0.473}}$ & 0.330   & 0.342  & 0.720   & $0.464^{\textcolor{red}{\downarrow 0.491}}$  \\
\multicolumn{1}{c|}{}                                          & Chimera-Medical & 0.300  & 0.300  & 1.000   & $0.462^{\textcolor{red}{\downarrow 0.445}}$ & 0.300  & 0.300  & 1.000   & $0.462^{\textcolor{red}{\downarrow 0.462}}$ & 0.300  & 0.300  & 1.000   & $0.462^{\textcolor{red}{\downarrow 0.472}}$ & 0.330   & 0.340  & 0.750   & $0.468^{\textcolor{red}{\downarrow 0.487}}$  \\ \hline
\end{tabular}
}
\end{table*}

% From the result, we found that XXX. Through in-depth analysis, we found that the reason is that. The result shows the effectiveness of the existing ITD methods and the usefulness of our collected \dataset.

% * Why do we evaluate DL-Based LAD while not for role-based methods?

% Effectiveness Evaluation of Existing ITD Methods

% 1. LSTM/GNN/IF for \cert dataset (code available)
% 2. Multiple attack eval, attack success rate

\subsection{Performance under Different Foundation Models}

We further investigate how different foundation models affect the effectiveness of \chimera. Concretely, we employ three models, including Gemini, GPT, and DeepSeek, to simulate a technology company and collect logs. Then, we quantify the quality of collected logs. 

\begin{figure}[t]
    \setlength{\abovecaptionskip}{0.05cm}
    \setlength{\belowcaptionskip}{0cm}
    \centering
    \begin{subfigure}{.8\linewidth}
        \includegraphics[width=\linewidth]{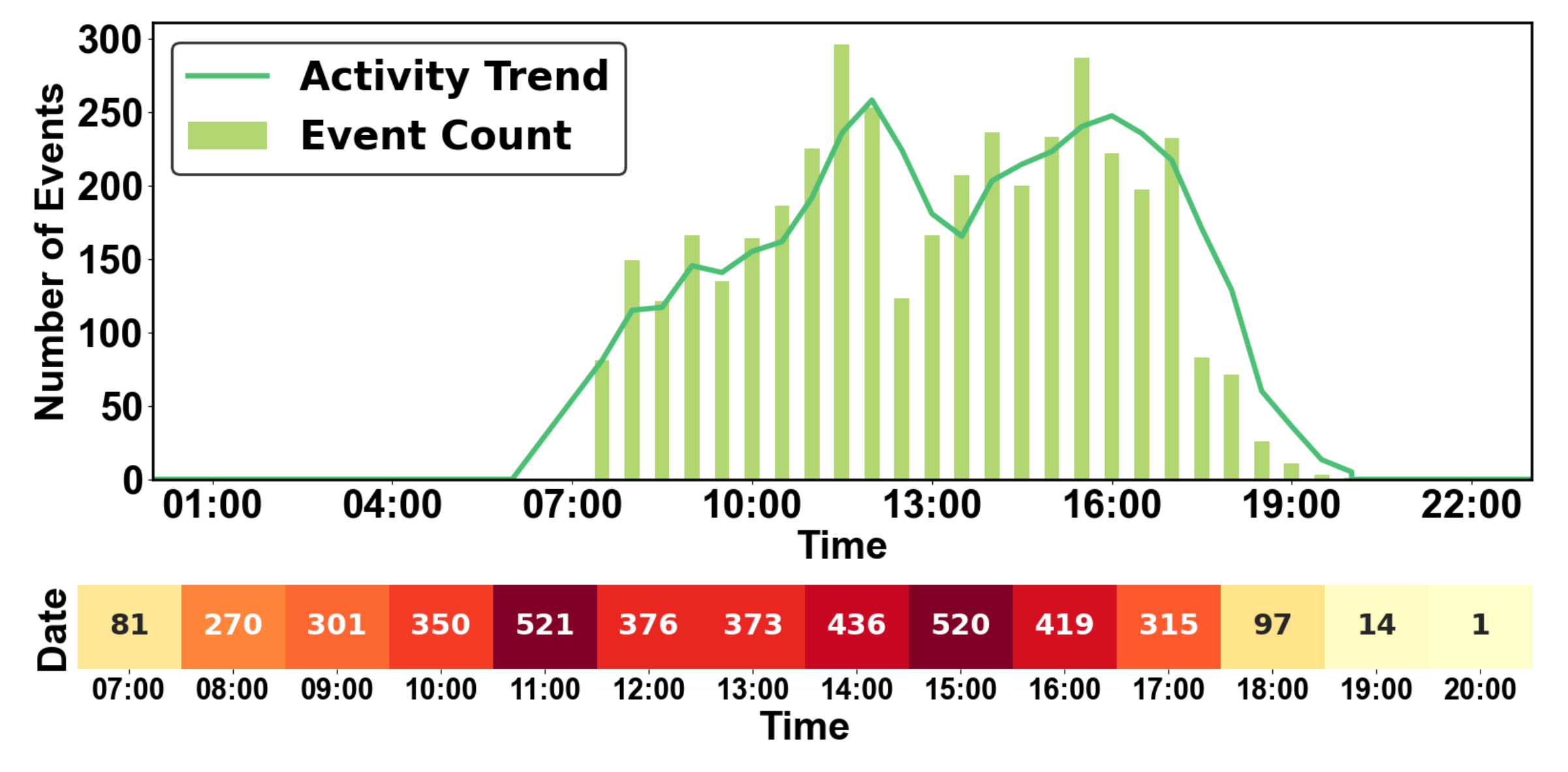}
        \caption{Daily benign employee activity based on OpenAI.}
        \label{fig:activity-openai}
    \end{subfigure}
    \begin{subfigure}{.8\linewidth}
        \includegraphics[width=\linewidth]{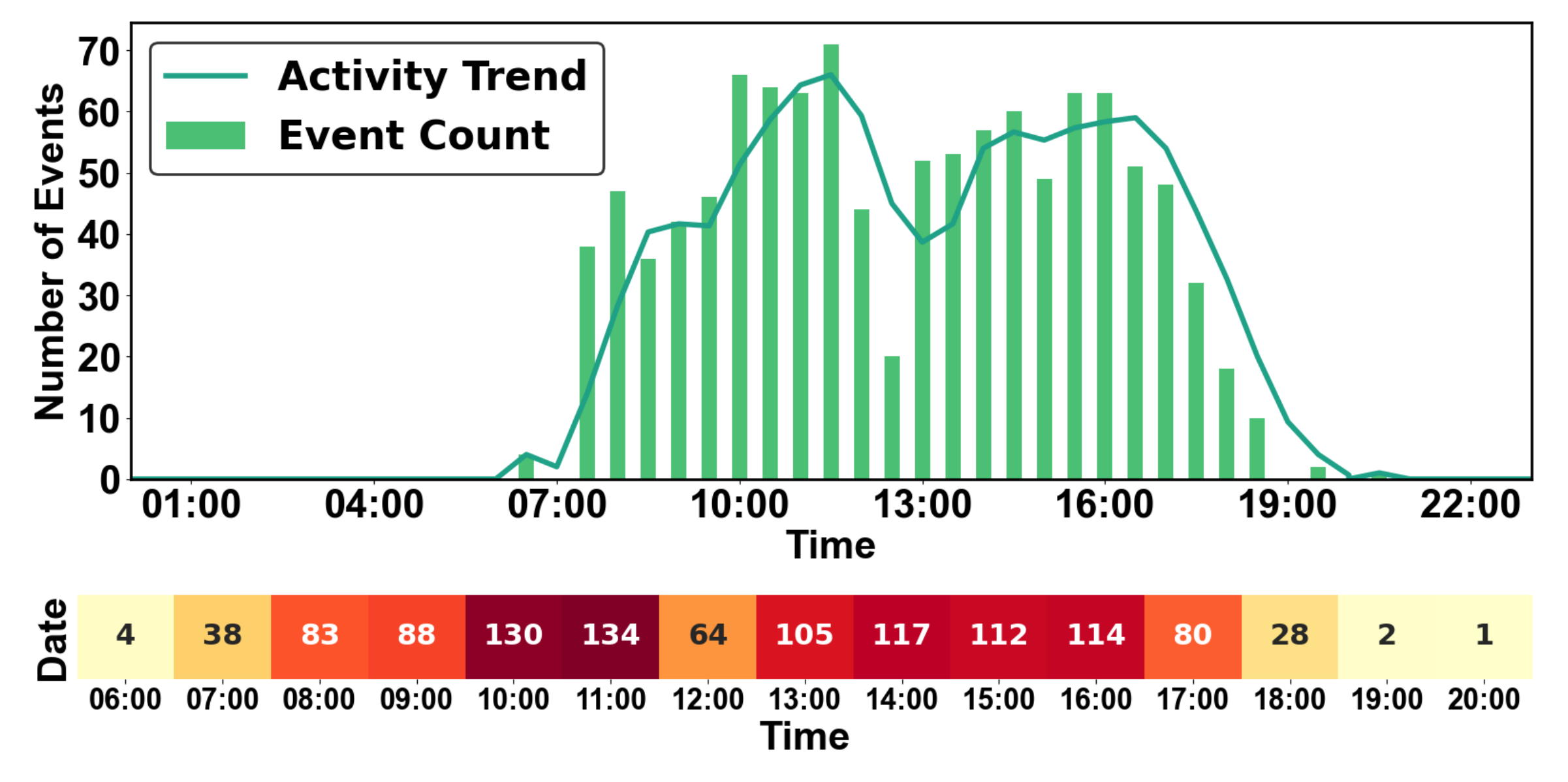}
        \caption{Daily benign employee activity based on Gemini.}
        \label{fig:activity-gemini}
    \end{subfigure}
    \begin{subfigure}{.8\linewidth}
        \includegraphics[width=\linewidth]{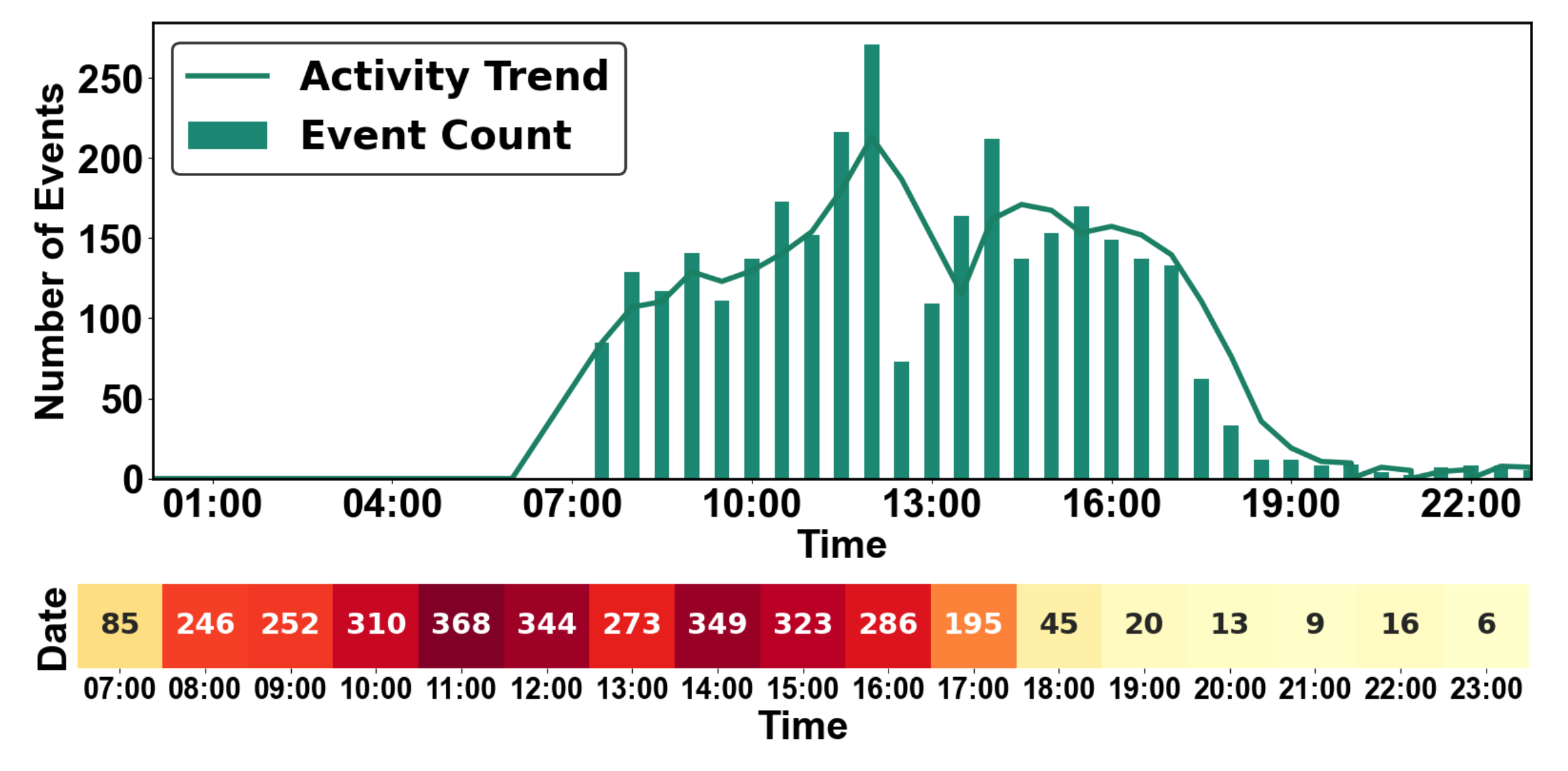}
        \caption{Daily benign employee activity based on DeepSeek.}
        \label{fig:activity-deepseek}
    \end{subfigure}
    \caption{Comparison of benign employee activities produced by \chimera with different foundation models.}
    \label{fig:activity-comparison-models}
\end{figure}

Table~\ref{tab:foundation_models} summarizes the statistical characteristics of datasets generated using different foundation models. Although all simulations use identical prompts, agent behaviors vary noticeably across models. In particular, agents powered by DeepSeek frequently extend work activities late into the night, sometimes until midnight, during tasks such as iterative document refinement. This behavior appears to stem from non-terminating reasoning loops, in which agents repeatedly attempt to refine content without reaching convergence, despite explicit constraints specified in the prompts. Such behavior is not observed in simulations conducted with GPT or Gemini.

Figure~\ref{fig:activity-comparison-models} illustrates the distribution of average daily activities over different foundation models. We observe that GPT facilitates the most diverse and communicative log outputs, producing higher volumes of events and richer insider interactions compared to Gemini and DeepSeek. In terms of task execution, \chimera performs well on atomized, agent-specific tasks, such as searching for API references or emailing colleagues, achieving nearly a 100\% success rate. Among the three, Gemini exhibits the highest failure rate on initial attempts, with 22.5\% of atomized tasks unresolved on the first try. Error analysis reveals that approximately 85\% of failures are due to incorrect tool usage (e.g., misformatted parameters or hallucinated function calls like browser\_tool or search\_google). Additional operational issues included API rate limits and occasional foundation model timeout errors.

\begin{table}[t]
\centering
\caption{Average statistics of the simulated dataset from different foundation models. \textit{DS} and \textit{DE} denote the average daily start and end times. \textit{TFR} denotes the task failure rate, and \textit{AS} denotes the number of attack steps.}
\label{tab:foundation_models}
\resizebox{.6\linewidth}{!}{
\begin{tabular}{|c|c|c|c|c|c|c|}
\hline
\textbf{Model} & \textbf{DS} & \textbf{DE} & \textbf{Event} & \textbf{Communication} & \textbf{TFR} & \textbf{AS} \\ \hline
\textbf{OpenAI} & 07:41 & 19:48 & 6627 & 2829 & 985 & 30 \\ \hline
\textbf{Gemini} & 06:58 & 19:01 & 4522 & 349 & 1017 & 25 \\ \hline
\textbf{DeepSeek} & 07:43 & 22:47 & 4735 & 390 & 872 & 26 \\ \hline
\end{tabular}
}
\end{table}

To investigate the usability of datasets generated by \chimera with different foundation models, we use them to build machine learning based ITD models and evaluate their effectiveness. Results shown in Table~\ref{tab:rq3-itd} indicate that logs from all three foundation models are effective for training detection models, yielding performance consistent with our prior evaluation findings.

\begin{table*}[t]
\setlength{\abovecaptionskip}{0.05cm}
\setlength{\belowcaptionskip}{0cm}
\centering
\caption{Evaluation results of ITD models in the dataset collected from different foundation models.}
\label{tab:rq3-itd}
\resizebox{\linewidth}{!}{
\begin{tabular}{c|cccc|cccc|cccc|cccc}
\hline
\multirow{2}{*}{\diagbox{\textbf{Baseline}}{\textbf{Dataset}}} & \multicolumn{4}{c|}{\textbf{ChimeraLog-OpenAI}}  & \multicolumn{4}{c|}{\textbf{ChimeraLog-Gemini}}  & \multicolumn{4}{c|}{\textbf{ChimeraLog-DeepSeek}} & \multicolumn{4}{c}{\textbf{CERT}}                \\ \cline{2-17} 
                                 & Acc   & \multicolumn{1}{l}{Pre} & Recall & F1    & Acc   & \multicolumn{1}{l}{Pre} & Recall & F1    & Acc    & \multicolumn{1}{l}{Pre} & Recall & F1    & Acc   & \multicolumn{1}{l}{Pre} & Recall & F1    \\ \hline
\textbf{SVM}                     & 0.758 & 0.68                    & 0.831  & 0.748 & 0.792 & 0.647                   & 0.781  & 0.708 & 0.798  & 0.724                   & 0.85   & 0.782 & 0.916 & 0.890                   & 0.937  & 0.913 \\ \hline
\textbf{CNN}                     & 0.859 & 0.889                   & 0.748  & 0.812 & 0.813 & 0.864                   & 0.735  & 0.794 & 0.825  & 0.893                   & 0.766  & 0.825 & 0.921 & 0.900                   & 0.937  & 0.918 \\ \hline
\textbf{GCN}                     & 0.693 & 0.669                   & 0.735  & 0.700 & 0.731 & 0.709                   & 0.755  & 0.731 & 0.736  & 0.704                   & 0.75   & 0.726 & 0.918 & 0.894                   & 0.937  & 0.915 \\ \hline
\textbf{DS-IID}                  & 0.834 & 0.732                   & 0.948  & 0.826 & 0.801 & 0.724                   & 0.902  & 0.803 & 0.863  & 0.695                   & 0.948  & 0.802 & 0.973 & 0.790                   & 0.860  & 0.824 \\ \hline
\end{tabular}
}
\end{table*}

\begin{tcolorbox}[size=title,opacityfill=0.1,breakable]

\textbf{Finding 3:} The choice of deployed foundation model affects the quality of generated data by \chimera, where GPT-4o serves as the best model. It is a trade-off to consider the data quality and budget.

% For Gemini, OpenAI, and Deepseek, under the same design of \chimera, what is the activity distribution and what patterns exist.    
\end{tcolorbox}

% Discussion
\section{Discussion}

\subsection{Implications} Based on our evaluation results and findings, we identify several key implications that could serve as promising guidance for future research.

\begin{itemize}[leftmargin=*]
    \item \textbf{Promising Usage of Multi-Agents.} Many works~\cite{li2023camel,wu2023autogen} employ LLM-based multi-agents to solve concrete tasks, such as software development~\cite{hong2023metagpt}. In addition, one great use of multi-agents should be simulating scenarios and constructing datasets that are difficult to collect manually, similar to \chimera.

    \item \added[id=R3]{\textbf{Distribution Shift Matters.} System updates and evolving employee behaviors cause distribution shifts~\cite{hu2022empirical} that can significantly degrade ITD model performance. While static datasets cannot reflect such evolution, \chimera can automatically regenerate realistic logs under updated configurations, enabling continual retraining and evaluation. By supporting dataset renewal aligned with real operational changes, real-world practitioners can monitor and mitigate performance decline caused by shifting data distributions.}
    
    \item \textbf{Towards Fully Automation with Advanced Simulation} The success of \chimera stems from LLM capabilities, yet it still falls short of realistic and intelligent simulation. Building threat environments and attack workflows remains manual. We urgently need advanced LLM-powered automation, such as threat-environment generation, personality-aware agent simulation, and autonomous penetration testing, to reduce manual work and enable intelligent simulations.
    
\end{itemize}

\subsection{Future Work}
\begin{itemize}[leftmargin=*]
\item \textbf{Cognitive and Hierarchical Realism.}  
Future integration of \chimera will enhance realism at both agent and organizational levels. At the agent level, \chimera can integrate personality modeling, short- and long-term memory, and psychological factors such as motivation and risk tolerance to enable more consistent human-like behaviors. At the organizational level, \chimera can simulate hierarchical enterprises with departments, branches, and subsidiaries, supporting cross-team interactions and exposing more realistic attack surfaces.

\item \textbf{Autonomous Red-Teaming and Adaptive Threat Evolution.} While current attack profiles are partially predefined, future work will enable more automated adversarial agents to strategically design and evolve multi-stage attacks. These agents can adapt their strategies, such as staged privilege escalation, based on contextual cues and defender behavior. This capability would transform \chimera into a generative environment for realistic and adaptive threat simulation. With longer simulation horizons spanning months or years, it would further support the study of slow-burn campaigns, including Advanced Persistent Threats (APTs), and attacker adaptation under distribution shifts.

\item \textbf{Co-Evolutionary Defense and Human-in-the-Loop Simulation.} To systematically explore mitigation strategies, \chimera can incorporate defensive agents and interactive feedback from security experts during simulation. This human-in-the-loop design will allow analysts to review generated activities, adjust agent objectives, or participate directly as attackers or defenders, enabling the simulation to self-correct and improve behavioral plausibility. By integrating adaptive defense mechanisms and expert guidance, \chimera will support long-term co-evolution studies of attacker–defender dynamics and the continual retraining of ITD models under realistic operational drift.
\end{itemize}

\subsection{Ethics and Societal Impact}
\label{sec:ethics}
\chimera is designed solely for academic research on insider-threat detection, focusing on malicious insiders who engage in behaviors such as intellectual-property theft, data exfiltration, sabotage, or unauthorized access for personal or corporate gain. The framework does not model politically motivated scenarios and analyzes all abnormal activities strictly from an organizational risk perspective.

\begin{itemize}[leftmargin=*]
\item \added[id=R3]{\textbf{System-Level Isolation and Compromisability.} To ensure safe and controlled simulations~\cite{li2026webcloak,luo2025agentauditor}, all agent activities should be executed inside dedicated containers with restricted privileges, effectively sandboxing every process from the host system. 
Each container operates in an isolated virtual network segment and communicates only through well-defined internal channels. The logging runs at a higher level than the agents and mounts logs in append-only mode, preventing any modification by simulated users.}

\item \added[id=R3]{\textbf{External Communication Safeguards.} 
For simulation functionality requiring Internet access, such as website browsing or online form submitting, \chimera employs internal mock web servers that emulate realistic responses without contacting the real Internet. 
Outbound traffic is controlled through a strict whitelist-based access policy, configurable per simulated organization, which restricts all network interactions to approved internal domains.}

\item \textbf{Privacy and Synthetic Data Integrity.} 
All data generated in \dataset are fully synthetic, including user names, email addresses, and IPs. No real-world personal or organizational information is used. 
To further mitigate the potential risk of inadvertent data leakage through LLM generations~\cite{chu2024reconstruct}, we apply vocabulary mutation and normalization to all generated text before storage. Consequently, \chimera contains no identifiable personal information while maintaining the semantic and structural realism necessary for insider-threat research.

\item \textbf{Pre-Screening of Attack Scenarios.} 
All attack profiles and corresponding prompts are manually reviewed to ensure that no action can affect real entities, download live malware, or propagate harmful content. The resulting dataset, therefore, represents safe, ethically bounded simulations suitable for reproducible academic research.
\end{itemize}

% Conclusion
\section{Conclusion}

In this paper, we introduced a multi-agent-based framework \chimera, which is designed for simulating internal corporate activities in enterprise environments. Based on \chimera, we constructed a new dataset \dataset that contains diverse internal operation logs to support evaluating ITD methods. Human studies demonstrated that \dataset is as realistic as real-world insider threat datasets. Experiments on four ITD methods showed that \dataset is more challenging than existing datasets, and distribution shifts posed significant concerns for ITD methods. Based on these findings, we summarized several implications from different perspectives. We believe this work can, to some extent, facilitate future research in enhancing enterprise security.

\bibliographystyle{unsrtnat}
\bibliography{references}  %%% Uncomment this line and comment out the ``thebibliography'' section below to use the external .bib file (using bibtex) .




% 	\bibitem{keshet2016prediction}
% 	Keshet, Renato, Alina Maor, and George Kour.
% 	\newblock Prediction-Based, Prioritized Market-Share Insight Extraction.
% 	\newblock In {\em Advanced Data Mining and Applications (ADMA), 2016 12th International 
%                       Conference of}, pages 81--94,2016.

% \appendix
% \input{Chapters/Appendix}

% \end{thebibliography}

\end{document}